\documentclass[twocolumn]{aastex631}

\shorttitle{Bright Strongly Lensed Galaxies at $3 < z < 4$}
\shortauthors{Zhang, Manwadkar et al.}

\usepackage{appendix}
\usepackage{float}
\graphicspath{{./}{figures/}}

\renewcommand{\arraystretch}{1.5}

\begin{document}

\title{COOL-LAMPS. IV. A Sample of Bright Strongly Lensed Galaxies at $3 < z < 4$}

\author[0000-0001-6454-1699]{Yunchong Zhang} 
\affiliation{Department of Astronomy and Astrophysics, University of Chicago, 5640 South Ellis Avenue, Chicago, IL 60637, USA}
\author[0000-0002-7113-0262]{Viraj Manwadkar}
\affiliation{Department of Astronomy and Astrophysics, University of Chicago, 5640 South Ellis Avenue, Chicago, IL 60637, USA}
\affiliation{Department of Physics, Stanford University, 382 Via Pueblo Mall, Stanford, CA 94305, USA}
\affiliation{Kavli Institute for Particle Astrophysics \& Cosmology, P. O. Box 2450, Stanford University, Stanford, CA 94305, USA}
\author[0000-0003-1370-5010]{Michael D. Gladders}
\affiliation{Department of Astronomy and Astrophysics, University of Chicago, 5640 South Ellis Avenue, Chicago, IL 60637, USA}
\affiliation{Kavli Institute for Cosmological Physics, University of Chicago, 5640 South Ellis Avenue, Chicago, IL 60637, USA}
\author[0000-0002-3475-7648]{Gourav Khullar}
\affiliation{Department of Astronomy and Astrophysics, University of Chicago, 5640 South Ellis Avenue, Chicago, IL 60637, USA}
\affiliation{Kavli Institute for Cosmological Physics, University of Chicago, 5640 South Ellis Avenue, Chicago, IL 60637, USA}
\affiliation{Department of Physics and Astronomy and PITT PACC, University of Pittsburgh, Pittsburgh, PA 15260, USA}

\author[0000-0003-2200-5606]{H{\aa}kon Dahle}
\affiliation{Institute of Theoretical Astrophysics, University of Oslo, P.O. Box 1029, Blindern, NO-0315 Oslo, Norway}
\author[0000-0003-4470-1696]{Kate A. Napier}
\affiliation{Department of Astronomy, University of Michigan, 1085 S. University Ave, Ann Arbor, MI 48109, USA}
\author[0000-0003-3266-2001]{Guillaume Mahler}
\affiliation{Centre for Extragalactic Astronomy, Durham University, South Road, Durham DH1 3LE, UK}
\affiliation{Institute for Computational Cosmology, Durham University, South Road, Durham DH1 3LE, UK}
\author[0000-0002-7559-0864]{Keren Sharon}
\affiliation{Department of Astronomy, University of Michigan, 1085 S. University Ave, Ann Arbor, MI 48109, USA}

\author[0000-0001-9225-972X]{Owen S. Matthews Acu\~{n}a}
\affiliation{Department of Astronomy and Astrophysics, University of
Chicago, 5640 South Ellis Avenue, Chicago, IL 60637, USA}
\author{Finian Ashmead}
\affiliation{Department of Astronomy and Astrophysics, University of Chicago, 5640 South Ellis Avenue, Chicago, IL 60637, USA}
\author[0000-0003-1697-7062]{William Cerny}
\affiliation{Department of Astronomy and Astrophysics, University of Chicago, 5640 South Ellis Avenue, Chicago, IL 60637, USA}

\author[0000-0002-7868-9827]{Juan Remolina Gonz\`{a}lez}
\affiliation{Department of Astronomy, University of Michigan, 1085 S. University Ave, Ann Arbor, MI 48109, USA}

\author[0000-0003-2294-4187]{Katya Gozman}
\affiliation{Department of Astronomy, University of Michigan, 1085 S. University Ave, Ann Arbor, MI 48109, USA}

\author[0000-0001-8000-1959]{Benjamin C. Levine}
\affiliation{Department of Astronomy and Astrophysics, University of Chicago, 5640 South Ellis Avenue, Chicago, IL 60637, USA}
\affiliation{Department of Physics and Astronomy, Stony Brook University, Stony Brook, NY 11794, USA}

\author{Daniel Marohnic}
\affiliation{Department of Astronomy and Astrophysics, University of Chicago, 5640 South Ellis Avenue, Chicago, IL 60637, USA}

\author[0000-0002-8397-8412]{Michael N. Martinez}
\affiliation{Department of Astronomy and Astrophysics, University of
Chicago, 5640 South Ellis Avenue, Chicago, IL 60637, USA}

\author[0000-0001-5931-5056]{Kaiya Merz}
\affiliation{Department of Astronomy and Astrophysics, University of
Chicago, 5640 South Ellis Avenue, Chicago, IL 60637, USA}

\author[0000-0002-7922-9726]{Yue Pan}
\affiliation{Department of Astronomy and Astrophysics, University of
Chicago, 5640 South Ellis Avenue, Chicago, IL 60637, USA}

\author[0000-0002-9142-6378]{Jorge A. Sanchez}
\affiliation{Department of Astronomy and Astrophysics, University of
Chicago, 5640 South Ellis Avenue, Chicago, IL 60637, USA}
\author{Isaac Sierra}
\affiliation{Department of Astronomy and Astrophysics, University of Chicago, 5640 South Ellis Avenue, Chicago, IL 60637, USA}
\author[0000-0002-2358-928X]{Emily E. Sisco}
\affiliation{Department of Astronomy and Astrophysics, University of
Chicago, 5640 South Ellis Avenue, Chicago, IL 60637, USA}
\author[0000-0002-1106-4881]{Ezra Sukay}
\affiliation{Department of Astronomy and Astrophysics, University of
Chicago, 5640 South Ellis Avenue, Chicago, IL 60637, USA}
\author[0000-0001-6584-6144]{Kiyan Tavangar}
\affiliation{Department of Astronomy and Astrophysics, University of
Chicago, 5640 South Ellis Avenue, Chicago, IL 60637, USA}

\author[0000-0002-6779-4277]{Erik Zaborowski}
\affiliation{Department of Astronomy and Astrophysics, University of Chicago, 5640 South Ellis Avenue, Chicago, IL 60637, USA}

\correspondingauthor{Yunchong Zhang}
\email{aruba19th@uchicago.edu}

\begin{abstract}
We report the discovery of five bright strong gravitationally lensed galaxies at $3 < z < 4$: COOL\,J0101$+$2055 ($z = 3.459$), COOL\,J0104$-$0757 ($z = 3.480$), COOL\,J0145$+$1018 ($z = 3.310$), COOL\,J0516$-$2208 ($z = 3.549$), and COOL\,J1356$+$0339 ($z = 3.753$). These galaxies have magnitudes of $r_{\rm AB}, z_{\rm AB} < 21.81$ mag and are lensed by galaxy clusters at $0.26 < z < 1$. This sample nearly doubles the number of known bright lensed galaxies with extended arcs at $3 < z < 4$.  We characterize the lensed galaxies using ground-based {\it grz}/{\it giy} imaging and optical spectroscopy. We report model-based magnitudes and derive stellar masses, dust content, and star-formation rates via stellar population synthesis modeling. Building lens models based on ground-based imaging, we estimate source magnifications in the range $\sim$29 to $\sim$180. Combining these analyses, we derive demagnified stellar masses in the range $\rm log_{10}(M_{*}/M_{\odot}) \sim 9.69 - 10.75$ and star formation rates in the youngest age bin ranging from $\rm log_{10}(SFR/(M_{\odot}\cdot yr^{-1})) \sim 0.39 - 1.46$, placing the sample galaxies on the massive end of the star-forming main sequence in this redshift interval. In addition, three of the five galaxies have strong Ly$\alpha$ emissions, offering unique opportunities to study Ly$\alpha$ emitters at high redshift in future work.
\end{abstract}


\section{Introduction} 
\label{sec:intro}

Strongly lensed star-forming galaxies at high redshift provide unique opportunities to understand star formation processes in the early universe. Using the spectral energy distribution \citep[SED; see][for a review of this method]{Conroy.etal.2013} we can infer important physical properties of the galaxies such as the stellar mass ($M_{*}$), star-formation rate (SFR), and dust extinction \citep[e.g.,][]{Kennicutt.etal.1998,Calzetti.etal.2000,Giavalisco.etal.2002,Salim.etal.2007,Maraston.etal.2010,Walcher.etal.2011,Wilkins.etal.2012}. Strong lensing allows us to probe much smaller subgalactic spatial scales that would normally be inaccessible without the extra magnification that lensing provides. 

\begin{figure*}[h!tb]
    \centering
    \includegraphics[width = \textwidth]{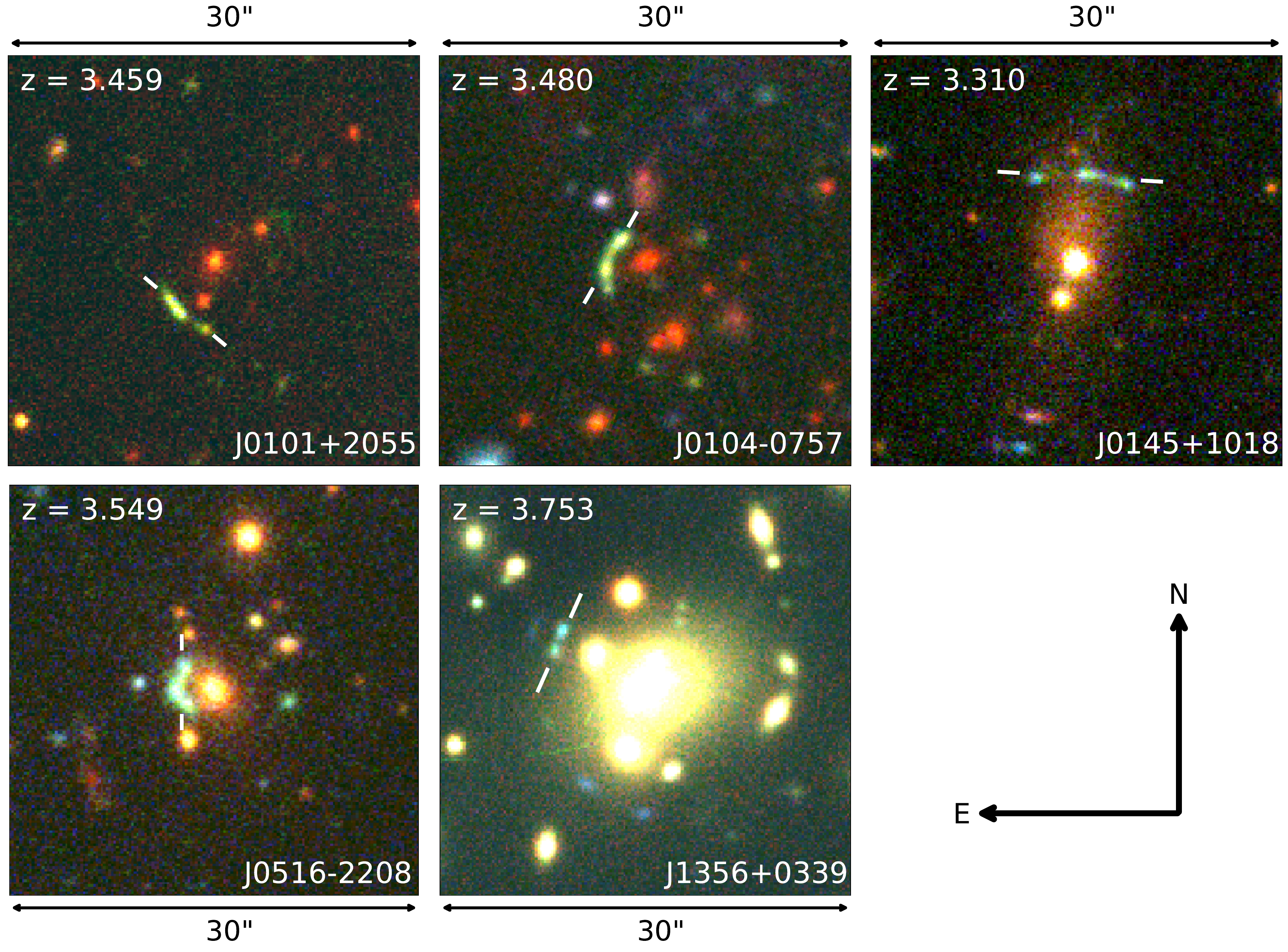}
    \caption{ RGB color images of size 30'' x 30''  of the lensed galaxies in our sample. The data for CJ0101, CJ0104, CJ0516, and CJ0145 are from Magellan LDSS3 \textit{grz} imaging. The data for CJ1356 are from the HSC-SSP \textit{giy} DR3 data release. The white ticks in each panel mark the lensed images that are confirmed spectroscopically. The spectroscopically confirmed redshift of the lensed galaxies is shown in the upper left corner of each panel. Note that the color contrast for each of these images has been enhanced to clearly show the lensed arcs. }
    \label{fig:lens_image}
\end{figure*}

Selecting strongly lensed galaxies consists, to first order, of two methods: deeply detailed studies of large lensing cross-section lenses, or mining large areas of sky across a wide range of foreground lens cross-sections to find the brightest examples. 
The former allows one to detect galaxies well below the detection limits reached in deep-field images and significantly improves the signal-to-noise ratio (S/N) of observations compared to their unlensed counterparts. Pointed (i.e., pencil-beam) strong lensing surveys such as the Hubble Frontier Fields \citep{Lotz.etal.2017} have enabled studies \citep[e.g.,][]{Santini.etal.2017,inaldi.etal.2022} to include fainter galaxies to probe the low stellar mass regime ($\log_{10}(M_{*}/M_{\odot})\lesssim 9.0 $) \citep[e.g.,][]{alavi2016}.
The latter approach has yielded exquisitely bright lensed sources where the magnification of strong lensing provides extreme spatial resolutions of only tens of parsecs in the source plane \citep[e.g.,][]{johnson2017}. A particularly cogent example of the benefits of such an object is the Sunburst Arc \citep{Dahle.etal.2016} where lensing allows identification of a single star-forming region that is leaking Lyman continuum (LyC) radiation, enabling unique opportunities to study the detailed physics of LyC escape
\citep[e.g.,][]{Rivera-Thorsen.etal.2017, Furtak.etal.2022}.

In the redshift range $3 < z < 4$, only six bright lensed star-forming galaxies with extended arcs have been published to date \citep[][]{Smail.etal.2007.cosmiceye,gilbank08,Oguri.etal.2012.sgas1050,Rigby.etal.2018.megasaura,Dessauges-Zavadsky17,Christensen12}.  This small sample has enabled a wide range of detailed observations and attendant analyses that would be difficult or impossible absent strong lensing. These include long-wavelength observations of molecular lines \citep{Coppin07}, resolved internal dynamics \citep{jones10}, high-S/N moderate resolution spectra at rest-UV wavelengths \citep{quider10,Bayliss.etal.2014.sgas1050,Rigby.etal.2018.megasaura}, and comparisons of said spectra to detailed models \citep{byler20}. 
In this work, we present the discovery and primary analysis of five additional UV-bright strongly lensed galaxies in the range $3 < z < 4$, nearly doubling the number of known objects in the same category. This redshift range is well before the peak of the cosmic SFR density at $z\sim2$ \citep[][]{madaureview2014} and the corresponding peak of the redshift distribution of prior bright lensed galaxy samples \citep{matt2011,matt2012,Stark.etal.2013.cassowary,Tran.etal.2022}; z=3.5 is approximately the half-time between the plausible beginning of star formation at $z\sim20$ and that later peak.



The goal of this paper is to determine the physical and evolutionary properties of these lensed galaxies by means of their photometric and UV spectral features. The structure of this paper is as follows. In Section \ref{sec:discovery}, we introduce the discovery of objects in this sample. Section \ref{sec:followup_obs} describes the photometric and spectroscopic data acquisition following the discoveries. In Section \ref{analysis}, we present the results from photometric modeling, lens modeling, and stellar population synthesis with photometric information from rest-frame UV bands. Section \ref{sec:discuss} discusses the limitation of our analysis and compares the initial results with other objects in the same regime. 


\begin{table}
\caption{The names and positions of our sample galaxies. Note that we name each lensing system by taking the position of the lensing galaxy. For COOL\,J0104$-$0757, we choose the position of the galaxy in the North at $z\simeq 1$ as the notional coordinate of the lensing system. The relative positions of lensed images are shown in Figure \ref{fig:lens_image}. As a shorthand, we will coalesce the system names to the four first digits of their coordinates (e.g., COOL\,J0101$+$2055 becomes CJ0101).}
    \centering
    \begin{tabular}{c|ccc}
    \hline
    \hline
     & RA (deg)& DEC (deg)   \\
     \hline
     {\rm COOL\,J0101$+$2055}  &  15.437643 & 20.928818  \\
      {\rm COOL\,J0104$-$0757} & 16.220170 & -7.952055  \\
      {\rm COOL\,J0145$+$1018} & 26.276362 & 10.310288 \\
      {\rm COOL\,J0516$-$2208} & 79.013054 & -22.146431 \\
      {\rm COOL\,J1356$+$0339} & 209.094447& 3.652319 \\
      \hline
    \end{tabular}
    \label{tab:basic}
\end{table}


In the entire analysis, we assume the cosmological parameters derived in  \citet{Planck.etal.2020}. All magnitudes are given in the AB system.  For inferred parameters with uncertainties, we report 16th, 50th, and 84th percentile values, unless otherwise specified. 

\section{Discovery}
\label{sec:discovery}
\subsection{Lens candidate selection}
The ChicagO Optically selected strong Lenses - Located At the Margins of Public Surveys (COOL-LAMPS) project began as a collaboration in an undergraduate research class and is an effort to find strong gravitational lenses in recent public imaging data. The lensing search is designed to find a wide variety of lenses and lensed sources \citep{Khullar.etal.2021,Sukay.etal.2022,COOLLAMPS3} but the targeted follow-up is primarily focused on lensed sources that are photometrically at the margins of the distributions of source color and brightness. The details and results of this search will be presented in an upcoming publication (COOL-LAMPS Collaboration in prep.). We select luminous red galaxies -- the most massive galaxies at a given epoch -- as potential sites of lensing by making cuts in color-magnitude diagrams of objects morphologically tagged as galaxies. For each selected line of sight, we construct a custom $grz$ color image, tuned to emphasize faint extended sources, with the cutout size scaled by an estimate of the galaxy richness at the potential lens redshift, estimated from a red-sequence early-type galaxy model. Each such image is scored from 0 (no lensing) to 3 (obvious unambiguous lensing) by multiple visual examiners, and a second pass is made through the resulting candidate list to cull objects for further consideration. The bulk of the lens selection work completed by the COOL-LAMPS collaboration so far has used the Dark Energy Camera Legacy Survey \citep[DECaLS;][]{Dey.etal.2019} datasets and all five lensing systems discussed here are taken from those data. The names and locations of our sample galaxies are given in Table~\ref{tab:basic}. The lensed sources listed here were selected as candidates for further follow-up due to high overall initial ranking from the lens search coupled with blue $r-z$ colors and redder $g-r$ colors suggestive of $z>3$ galaxies. 

\subsection{Spectroscopic confirmation}

CJ0101 and CJ0104 were observed spectroscopically using the 2.5m Nordic Optical Telescope and ALFOSC spectrograph on 2020 October 14. The total exposure time was 3000 s for CJ0101 and 2400 s for CJ0104. In both cases, we used grism 4 and a slit width of 1."3, producing a spectral resolution of R$\sim$300 and covering the wavelength range 3200-9600\AA. Those data measured initial source redshifts for both systems. On 2021 July 20, CJ0145 and CJ1356 were first observed spectroscopically for 1200 s each using the 1".0 slit in echelle mode of the FIRE near-IR spectrograph \citep{simcoe2013} on the Magellan I 6.5m Baade telescope; those data provided source redshifts from multiple emission lines in both instances. On the same night and telescope, CJ0516 was observed using the IMACS optical spectrograph \citep{Alan2011} for 2000 s in total using the f/4 camera, the 150 $lines\ mm^{-1}$ grating, and a 1".0 slit; those data also provided a source redshift from multiple emission and absorption features.




\section{Follow-up observations and redshifts}
\label{sec:followup_obs}

\begin{figure*}[t]
\centering
    \includegraphics[width=\linewidth]{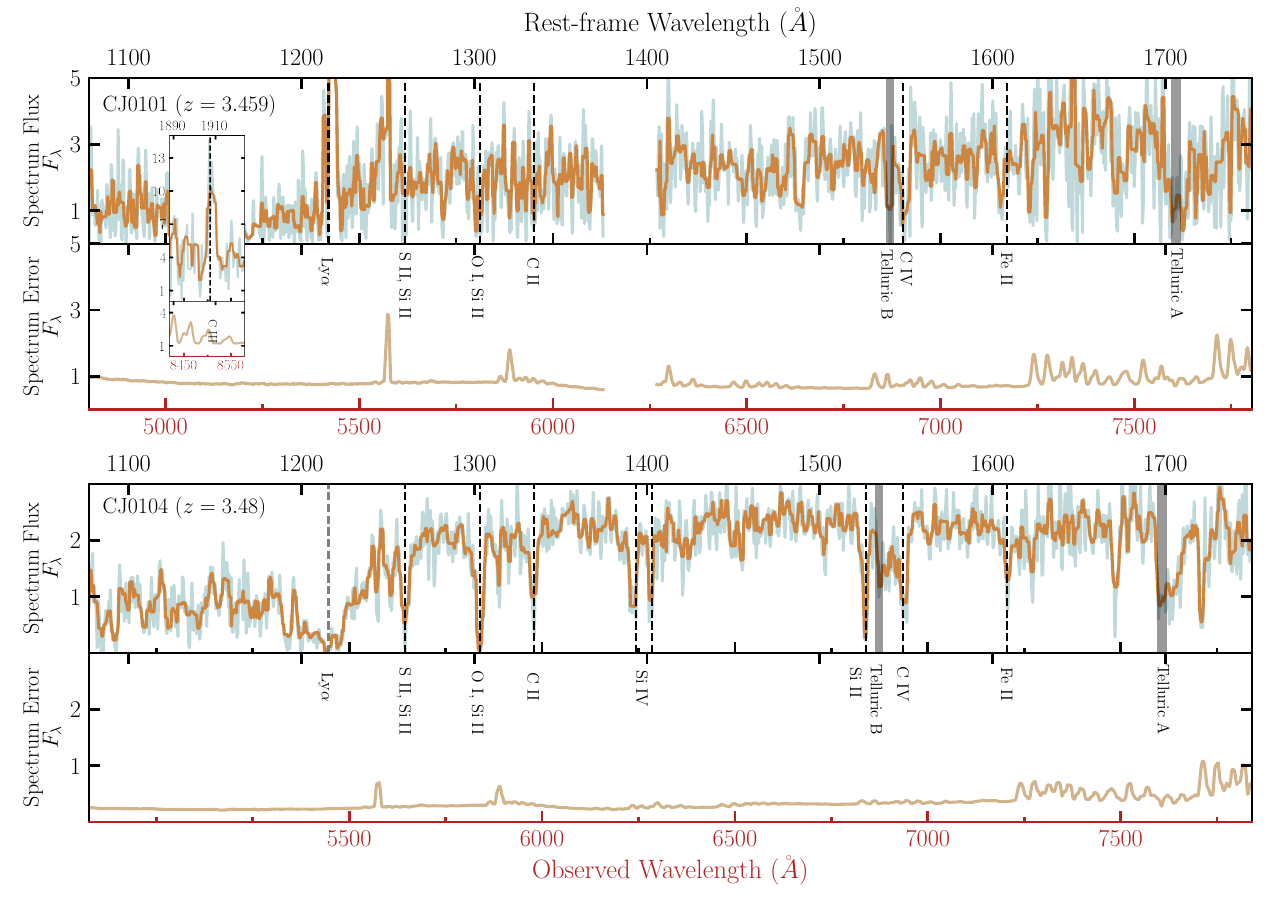}

\caption{Optical/near-IR spectra of the source galaxies of CJ0101 (Magellan/IMACS) and CJ0104 (Magellan/LDSS3). \textit{Top panels in each row}: the extracted raw spectrum (light blue) and the spectrum convolved with a median filter (orange); \textit{Bottom panels in each row}: the noise spectrum. The fluxes are in units of $\rm erg/s/ cm^2/ \AA$ and normalized arbitrarily. For each row, the extracted spectra and noise spectrum have the same scaling for fluxes. Black vertical dashed lines indicate significant spectral features used to infer the redshifts of the source galaxies. Gray vertical shaded regions mark the  telluric absorption bands. For CJ0101, we mark the Ly$\alpha$ emission line with a black dashed line and for CJ0104, we mark Ly$\alpha$ in absorption at $\rm 1215 \AA$ with a gray dashed line for reference. In the case of CJ0101, the emission line of C III] $\rm \lambda\lambda$1907, 1909 is present and shown in the inset. Note that there are several significant absorption features that are likely due to intervening absorbers in all object spectra. There are no spectral data in the gap in the first row due to the chip gaps of the IMACS detector mosaic.
\label{spectrums1}}
\end{figure*}

\begin{figure*}[t]
\centering
\includegraphics[width=\linewidth]{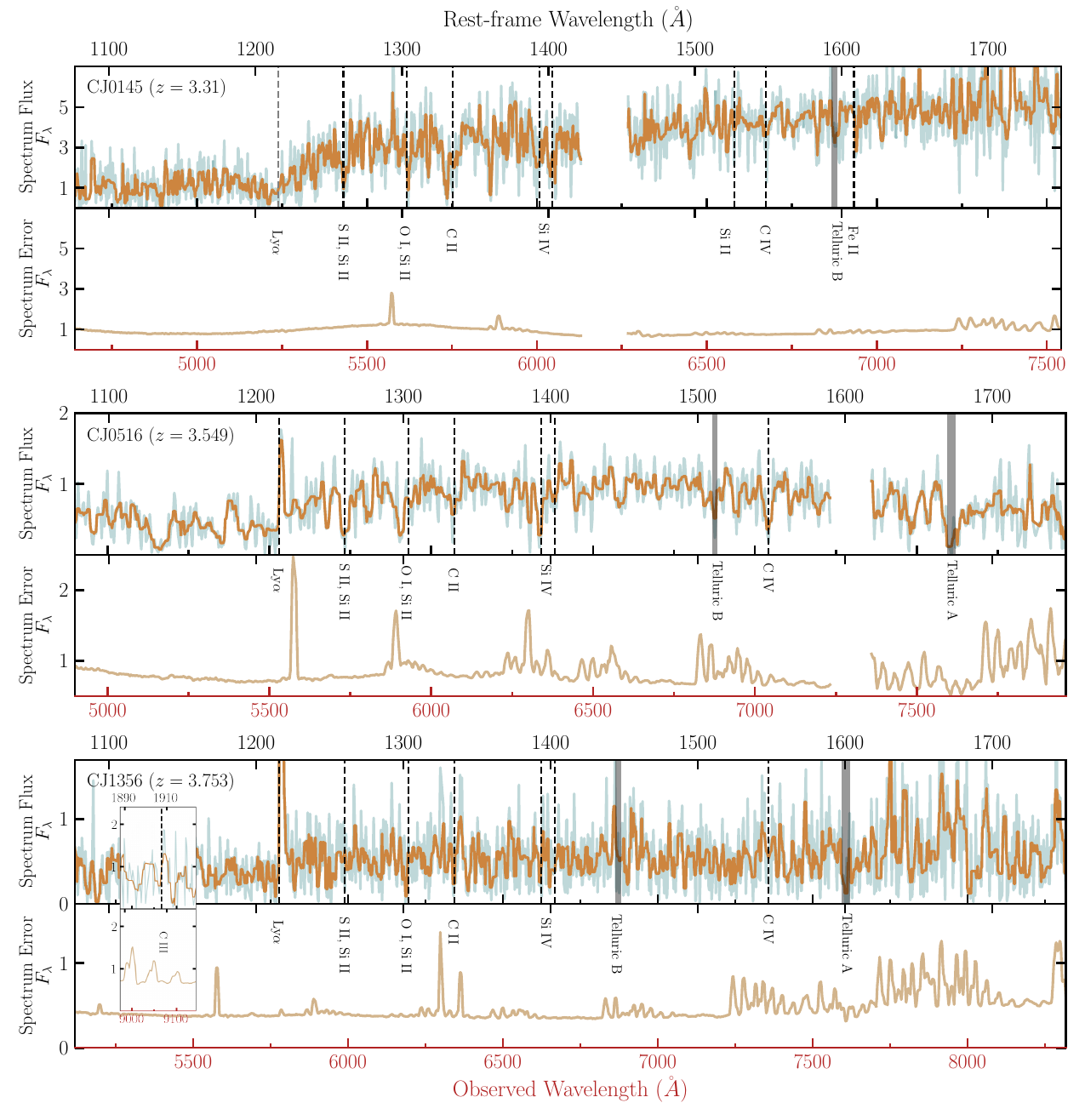}
\caption{Optical/near-IR spectra of the source galaxies of CJ0145 (Magellan/IMACS), CJ0516 (Magellan/IMACS) and CJ1356 (Magellan/LDSS3). \textit{Top panels in each row}: the extracted raw spectrum (light blue) and the spectrum convolved with a median filter (orange); \textit{Bottom panels in each row}: the noise spectrum. The fluxes are in units of $\rm erg/s/ cm^2/ \AA$ and normalized arbitrarily. For each row, the extracted spectra and noise spectrum have the same scaling for fluxes. Black vertical dashed lines indicate significant spectral features used to infer the redshifts of the source galaxies. Gray vertical shaded regions mark the  telluric absorption bands. We mark the Ly$\alpha$ emission line at $\rm 1215 \AA$ with black dashed lines in the cases of CJ0516 and CJ1356 and Ly$\alpha$ in absorption with a gray dashed line for reference in the case of CJ0145. For CJ1356, the emission line of C III] $\rm \lambda\lambda$1907, 1909 is present and shown in the inset. Note that there are several significant absorption features that are likely due to intervening absorbers in all object spectra. There are no spectral data in the gaps in the first and second rows due to the chip gaps of the IMACS detector mosaic.}
\label{spectrums2}
\end{figure*}

\subsection{Imaging}
\label{ssec:imaging}
We obtained broadband optical imaging, deeper and sharper than the discovery DECaLS data, in the \textit{grz} filters using the Magellan Clay telescope with the LDSS3C imaging spectrograph for four of the five lensed systems.
The exception, CJ1356, is present in the Hyper Suprime-Cam Subaru Strategic Program (HSC-SSP) DR3 data release, and we use the available \textit{giy} in lieu of additional Magellan data. The Magellan imaging data were acquired in clear conditions on 2020 August 13 and 2021 October 10 in seeing of 0."6 - 0."8, with total exposure times per filter ranging from 360 to 840 s selected by the source brightness and sky brightness at the time of observation.
Data were processed using standard techniques in IRAF \citep{Tody.1986,Tody.1993} and PHOTPIPE \citep{Rest.etal.2005,Garg.etal.2006,Miknaitis.etal.2007}, producing a stacked coadded image for each object and filter pair. An initial astrometric solution was defined using astrometry.net \citep{astrometry.net}; this was further refined using custom code to precisely match the astrometric solution for each image to the DECaLS catalogs. Photometric zero-points were derived by direct comparison to DECaLS and have, in most cases, uncertainties of less than 0.02 mag. The quoted uncertainty in the HSC-SSP DR3 data is 0.015 mag \citep{Aihara.etal.2021.DR3}.


Figure~\ref{fig:lens_image} shows color images, derived from the above data, for all five strong lensing systems. 


\subsection{Spectroscopy}
\label{ssec:spectra}

After these initial discovery observations, four of the five systems (excluding CJ0516) were reobserved spectroscopically using either the Magellan I 6.5m Baade telescope and IMACS or the Magellan II 6.5m Clay telescope and LDSS3C. The goal was to gather rest-frame UV spectra to characterize the Ly$\alpha$ line and, in most instances, place candidate secondary images or foreground lens galaxies on the same long slit to acquire further redshifts. Observations occurred on 2021 October 13-14 and 26 and 2022 July 22.  

The follow-up spectroscopy was processed with custom scripts using standard techniques to flatten, extract to 1D, and wavelength and flux calibrate the data, all implemented in IRAF. All objects with apparent spectra were extracted, calibrated, and analyzed, not always successfully, for redshift information.

\subsection{Redshifts}
\label{ssec:lens_redshifts}

The redshifts of sources and foreground lenses, as well as their uncertainties are tabulated in Table \ref{tab:redshift}. For lensed sources, we used a set of interstellar absorption features (S II $\rm \lambda$1260 + Si II $\rm \lambda$1260, O I $\rm \lambda$1302 + Si II $\rm \lambda$1304, C II $\rm \lambda$1335, Si IV $\rm \lambda\lambda$1394, 1403, and C IV $\rm \lambda\lambda$1548, 1551) to obtain a set of redshifts for each object. In addition, emission line C III] $\rm \lambda\lambda$1907, 1909 and absorption lines Si II $\rm \lambda$1527 and Fe II $\rm \lambda$1608 are present in the spectra of some objects. See Figure \ref{spectrums1} and \ref{spectrums2} for details. We report the final redshift values by taking the set average and the uncertainties by taking the standard deviation. Note that for CJ1356, the reported redshift refers to the doubly imaged green arcs in the northeast of the field, as marked by white bars in Figure \ref{fig:lens_image} and a second family of lensed images at $z \simeq 2.17$ is also present.

The redshifts of the lenses were obtained as follows. For CJ0104 and CJ0516, we extracted LDSS3/IMACS spectra of the brightest central galaxies (BCGs) in these systems. The BCG redshifts and related uncertainties, which were computed based on the spectral features, were taken as lens redshifts for these two systems. For CJ0101 and CJ1356, we used the BCG redshifts and uncertainties reported in eBOSS \citep{eBOSS}, which is part of the Sloan Digital Sky Survey (SDSS) DR16 \citep{SDSS_dr16}. For CJ0145, we obtained spectroscopic redshifts for two cluster members from our spectra and two other members from SDSS DR16. We computed the final redshift by taking the mean value of the cluster member redshifts and the uncertainties by taking the standard deviation.

\begin{table*}[thbp]
\caption{Table of Source and Lens Redshifts and Median Lens Magnification with 68\% Confidence Interval.}
    \centering
    \begin{tabular}{c|ccc}
    \hline
    \hline
     & $z_{\rm lens}$ & $z_{\rm source}$ & Magnification  \\
     \hline
     {\rm CJ0101}  &  $0.8708 \pm 0.0003 $ & $3.459 \pm 0.001$ & $100_{-63}^{+125}$ \\
      {\rm CJ0104} &  $1.0037 \pm 0.0006$ \& $0.8581 \pm 0.006$ & $3.480 \pm 0.002$& $40_{-16}^{+55}$\\
      {\rm CJ0145} &  $0.463 \pm 0.006$ & $3.310 \pm 0.003$ & $29_{-14}^{+23}$\\
      {\rm CJ0516} &  $0.658\pm 0.001 $ & $3.549 \pm 0.002$& $180_{-69}^{+173}$\\
      {\rm CJ1356} &  $0.26185 \pm 0.00006$& $3.753\pm 0.004$ & $93_{-19}^{+41}$\\
      \hline
    \end{tabular}
    \label{tab:redshift}
\end{table*}

\section{Analysis}\label{analysis}

\begin{figure*}[h]
    \centering
    \includegraphics[width = 0.72\textwidth]{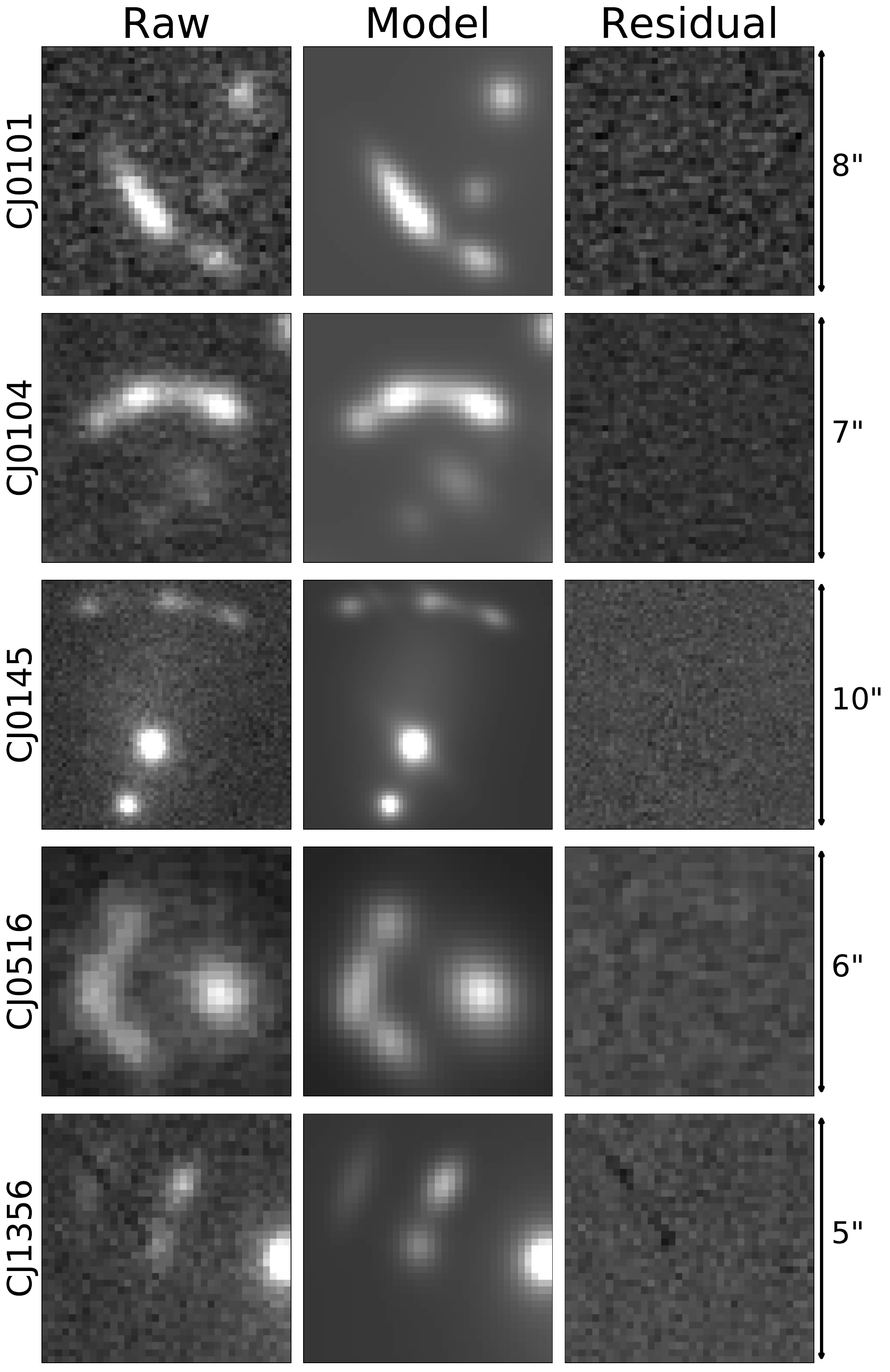}
    \caption{ Selected results of parametric photometry fitting: CJ1356 (HSC $g$ band), CJ0145 ($r$ band), CJ0516 ($r$ band), CJ0101 ($r$ band), and CJ0104 ($r$ band). The left column shows the original data frames, the central column shows the fitted model frames, and the right column shows the residual frames. The angular scale of each fitting frame is shown on the right in arcseconds for each row. Note that for CJ1356, the top left region of the image is contaminated by cosmic rays and has a series of pixels with underestimated values. Such a feature is not prominent in other band frames for this object. We manually identified the corrupted pixels and applied a mask for them during the fitting process. }
    \label{fig:galfit}
\end{figure*}

\subsection{Photometric Modeling with \texttt{GALFIT}}
\label{ssec:galfit}

To extract the photometry of the arc and relevant potential counterimages, we used the parametric model fitting code \texttt{GALFIT} \citep{Peng.etal.2010}. We measured the point-spread functions (PSFs) directly from the images by choosing bright, unsaturated, isolated point sources near the region of interest. We used multiple 2D Sérsic \citep{Sersic_1963} profiles to model each image. The model construction proceeded iteratively with single 2D Sérsic components being placed down on each source in the image and additional components added until the residuals were consistent with the background noise. We constructed the initial \texttt{GALFIT} models in the band where the lensed arc is the most prominent and then propagated that model with appropriate constraints to other bands. 

To estimate the statistical photometric uncertainties, we refitted the model multiple times by adding the final fitted \texttt{GALFIT} models to simulated instances of sky flux generated by bootstrapping the sky pixel distribution of the image. We obtained the sky pixels by iteratively clipping at 3$\sigma$ the pixel distribution around the image. The measured magnitude of the object in each of these refits gave us the uncertainty for each object. To compute the total magnitude uncertainty for each object, we combined the zero-point calibration uncertainty and above statistical uncertainties in quadrature. Reported object magnitudes were corrected for Galactic extinction. The LDSS3 imaging \textit{grz}-band magnitudes were corrected by values provided in the DECaLS database \citep{Dey.etal.2019} while the HSC imaging {\it giy}-band magnitudes were corrected by values reported in \citet{MKExt} as implemented by the NASA/IPAC Extragalactic Database’s extinction calculator\footnote{ \href{https://ned.ipac.caltech.edu/extinction_calculator}{The NASA/IPAC Extragalactic Database (NED)} is funded by the National Aeronautics and Space Administration and operated by the California Institute of Technology.}.

\begin{table*}[htpb]
\caption{The \texttt{GALFIT} Modeled Photometry Magnitudes (with 68\% confidence intervals) of These Galaxies. }
    \centering
    \begin{tabular}{c|cccccc}
    \hline
    \hline
     & \textit{g} & \textit{r} & \textit{z} & HSC \textit{g} & HSC \textit{i} & HSC \textit{y} \\
     \hline
     CJ0101  & $22.58^{+0.02}_{-0.04}$ & $21.29^{+0.03}_{-0.02}$ &  $20.89^{+0.05}_{-0.04}$ & - & - & - \\
      CJ0104 &  $21.87^{+0.03}_{-0.03}$ & $20.59^{+0.06}_{-0.05}$ & $20.31^{+0.04}_{-0.05}$ & - & - & - \\
      CJ0145 & $21.62^{+0.05}_{-0.05}$ & $20.81^{+0.04}_{-0.04}$ & $20.71^{+0.07}_{-0.09}$ & - & - & - \\
      CJ0516 &  $21.45^{+0.04}_{-0.03}$ & $20.40^{+0.02}_{-0.03}$ & $20.34^{+0.06}_{-0.05}$ & - & - & - \\
      CJ1356 &  - & - & - & $22.55^{+0.05}_{-0.06}$ & $21.69^{+0.08}_{-0.08}$ & $21.81^{+0.27}_{-0.26}$ \\
      \hline
    \end{tabular}
    {\footnotesize \tablecomments{These magnitudes are in the AB system with $grz$ imaging from the Magellan LDSS3 imager and $giy$-band imaging from Hyper Suprime-Camera (HSC). Magnitudes are for the bright arc images only (denoted by white bars in Figure \ref{fig:lens_image}) and magnitudes have been accounted for Galactic extinction.}}
    \label{tab:galfit_results}
\end{table*}

Table~\ref{tab:galfit_results} tabulates the \texttt{GALFIT}-measured photometric magnitudes and corresponding uncertainties for each arc. The $r$-band image (HSC $g$-band image for CJ1356), \texttt{GALFIT} model, and residuals of each object field are shown in Figure \ref{fig:galfit}.

\begin{figure*}[th]
    \centering
    \includegraphics[width = \textwidth]{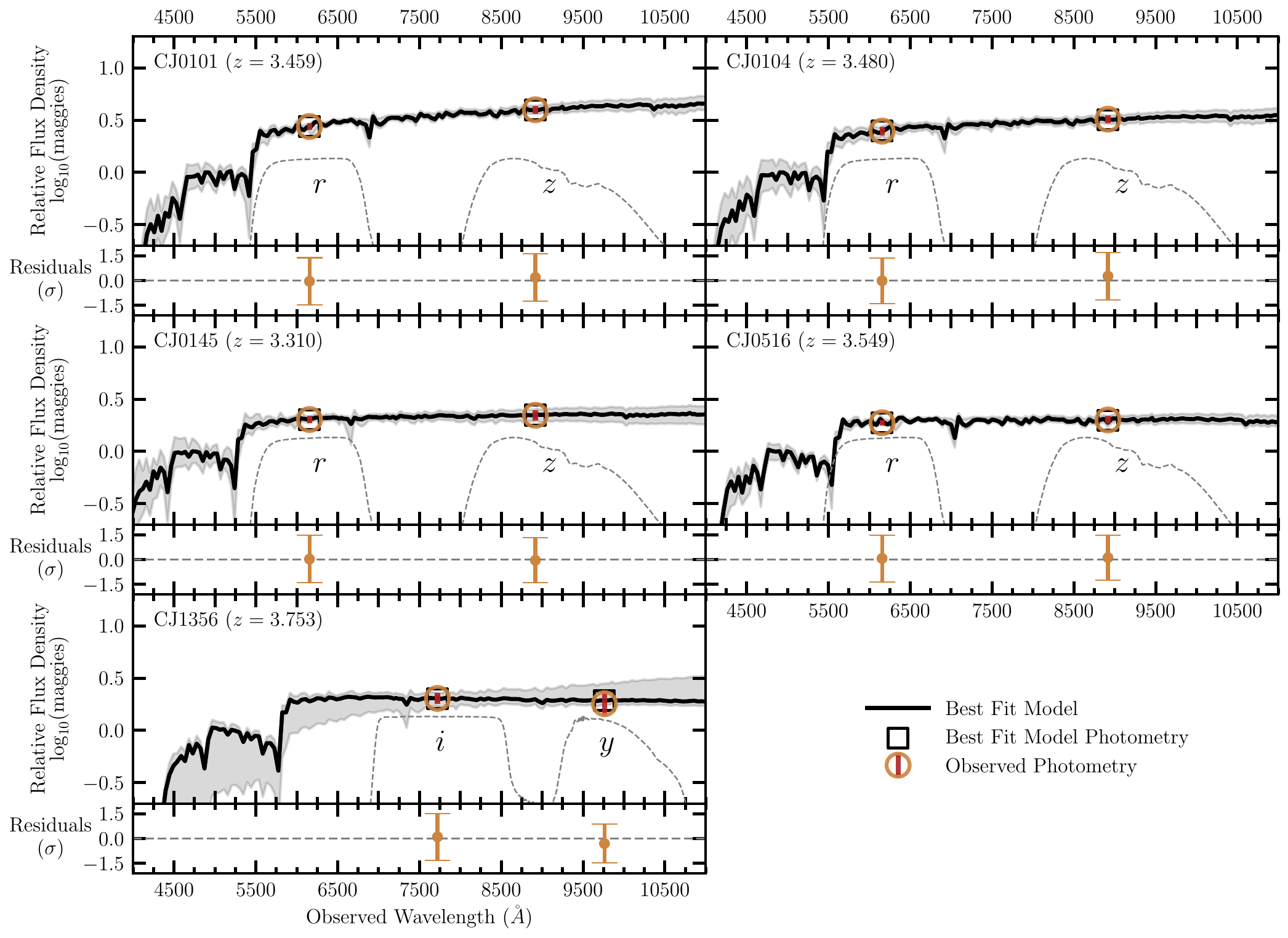}
    \caption{\textit{Top panels}: best fit fiducial SED models (black solid lines) for all five objects via \texttt{Prospector} using {\it rz/iy} photometry. The best-fit photometry is shown as black squares, while the observed photometry is shown as orange circles. The uncertainty of the observed photometry is represented by the length of the red bars in the orange circles. \textit{Bottom panels}: corresponding residual ($\chi$) values and errors. Note that everything is shown in observed wavelengths. The flux values of the entire model were renormalized by the flux value at $\rm \lambda = 1100 \AA$. The nebular line and continuum emissions were turned on during the fitting process. However, we chose not to display the nebular line emission in the model spectra.}
    \label{fig:all_seds1}
\end{figure*}

\newcommand{\sedcaption}{\texttt{Prospector} Analysis: Free Parameters in the Fiducial SED Models}
\begin{deluxetable*}{l cccc}
\tablecolumns{4}
\tablewidth{0pt}
\tablecaption{\sedcaption}
\tablehead{Parameter & Description & Priors}
\startdata
$M_{\star} (M_{\odot})$ & Total stellar mass formed & Log$_{10}$ Uniform: [10$^{9}$, 10$^{13}$] \\
$\log(Z/Z_{\odot})^{*}$ & Stellar metallicity & Uniform: [-2.0, 0.2]\\
$\tau_{\lambda, 2}$ & Diffuse dust optical depth & Tophat: [0.01, 3.00]\\
$\mathrm{gas}_{\log(Z/Z_{\odot})}^{*}$ & Gas-phase metallicity & Tophat: [-2, 0.2]\\
$\log(U)^{*}$ & Gas Ionization Parameter $U=n\gamma/n_H$ & Tophat: [-4.0, -1.5]\\
$\log(\mathrm{SFR}_{\rm ratio})$ & Ratio of the SFRs in adjacent age bins: & Student-T: mean=0.0,scale=0.3, $\nu=2$\\
& non-parametric SFH & Age bins: [0,50], [50,100], [100,500] and [500, age of Universe at z$_{obs}$]\\
\enddata
{\footnotesize \tablecomments{$*$ Considered as nuisance parameters.}}
\label{table:sed_model}
\end{deluxetable*}

\subsection{SED Fitting with \texttt{Prospector}}
\label{ssec:sed_fitting}

Using the model photometry in different filters (discussed in Section~\ref{ssec:galfit}), we performed Bayesian SED fitting for the lensed arc images. We used \texttt{Prospector}  \citep{Conroy.etal.2009,Conroy.etal.2010,Johnson.etal.2021.Prospector}, a stellar population synthesis and parameter inference framework based on Markov Chain Monte Carlo (MCMC). Note that the arc spectra only provide a spectroscopic redshift constraint and were not utilized in the SED fitting process; in this low-S/N regime, the spectra mostly sample saturated absorption lines from the interstellar medium, which are not  constrained by the \texttt{FSPS} libraries \citep{Conroy.etal.2009,Conroy.etal.2010}. Other than the slope of the rest-UV continuum, our spectral dataset does not provide additional information regarding the stellar population, which is already more reliably and robustly sampled by the photometry.

We chose to conduct SED fitting on the observed photometry rather than magnification-corrected photometry. We assumed that the magnification-dependent parameters in SED fitting (such as stellar mass and SFR) are not correlated with those that are not magnification-dependent (such as metallicity and dust extinction). This assumption has been tested and validated in previous COOL-LAMPS publications \citep{Khullar.etal.2021}.

We assumed a nonparametric star formation history (with age bins of $[0-50],[50-100],[100 -500]$ and $[500 - t_{\rm source}]$ Myr in look-back time). It is represented by a set of SFR$_{\rm ratio}$ values, which give the ratio of the SFRs in adjacent bins, and $t_{\rm source}$ is the age of the universe in megayears at the redshift of the source (see \citealt{Leja_2017} and \citealt{Khullar.etal.2021} for examples of implementation.) For example, for CJ0101, we have $t_{\rm source} = 1870$ Myr at $z=3.5$ assuming a \citet{Planck.etal.2020} cosmology. Throughout our entire analysis, we assume a Kroupa initial mass function \citep{Kroupa.etal.2001}. In the fiducial model, we fitted for free parameters such as total stellar mass formed in the galaxy $\log_{10}(M_{*}/M_{\odot})$, dust attenuation $\tau_{\lambda,2}$ (diffuse dust optical depth), stellar metallicity $\log_{10}(Z/Z_{\odot})$, gas ionization parameter and gas-phase metallicity. Refer to Table \ref{table:sed_model} for a list of these parameters and related priors.

For the five primary bright arcs photometered above, the $g$-band (or HSC $g$-band) photometry samples the Ly$\alpha$ line and/or Lyman forest. Since the emission and absorption of Ly$\alpha$ is complex and highly dependent on the gas geometry in the galaxies, the correlation between such processes and star-forming activity is not established in our sample galaxies \citep[e.g.,][]{calzetti1994, giavalisco1996, atek2008, sobral2019}. Hence, we chose to only use $rz$ photometry for the fiducial SED models.  The dust extinction and metallicities have flat and liberal priors, roughly covering the range allowed by the spectral model libraries. Gas ionization parameters and gas-phase metallicity have top-hat priors of $[-4,-1.5]$ and $[-2,0.2]$. We also included nebular continuum and line emission in our modeling process, while the line emission was turned off in the model spectrum.

\begin{figure*}[t]
    \centering
    \includegraphics[width = \linewidth]{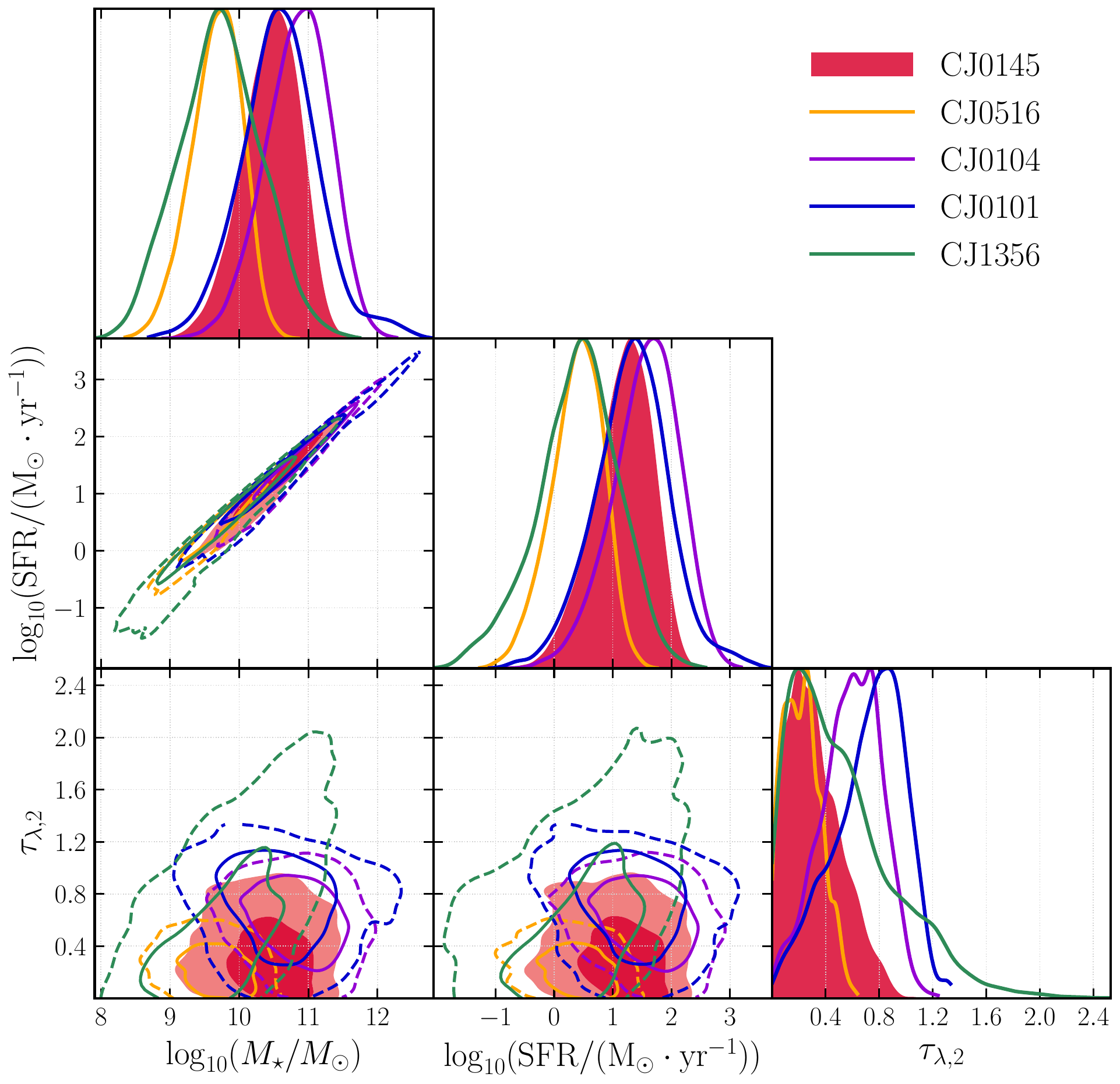}
    \caption{Corner plot with the posterior distributions and correlations for parameters from the \texttt{Prospector} SED fitting analysis for all the objects in this sample. The contours represent the 1$\sigma$(solid) and 2$\sigma$(dashed) confidence intervals. For visual clarity, one of the five systems is plotted with filled contours. Given the limitations of fitting models to a few data points, the measured stellar masses and SFRs span a significant range of values that significantly overlap. Refer to Section~\ref{ssec:sed_fitting} for details on the fitted parameters and Table~\ref{tab:sed_results} for the numerical values of these best-fit parameters.}
    \label{fig:corner}
\end{figure*}

The best-fit fiducial SED models are shown in Figure \ref{fig:all_seds1}. The residuals are shown in the lower panels for each fit to visually demonstrate the quality of the best-fit SEDs. In Figure \ref{fig:corner}, we present the posterior distributions of the best-fit parameters both as histograms and as pair-wise comparison via a corner plot. Refer to Table~\ref{tab:sed_results} for the best-fit values of the parameters.

Since we are only fitting for $rz$-band photometry in the fiducial model, our model explicitly captures the rest-frame UV properties of the galaxy spectrum, which is closely related to the emission from the hot and massive stellar populations and the dust extinction in the galaxy (e.g., see \citealt{salim2007, shivaei2015}, and references therein). Meanwhile, due to a lack of information input regarding the cooler and less massive stellar populations, properties such as stellar and gas-phase metallicity are not constrained in the model.  Hence, in Figure \ref{fig:corner}, we only show the free parameters that are constrained by the rest-frame UV observations, which are the total stellar mass $M_{*}$, instantaneous and dust-unobscured SFR (defined as the average SFR in the age bin closest to the epoch of observation, SFR$_{\rm last-bin}$) and dust attenuation $\tau_{\lambda,2}$. Note that the stellar mass and SFR shown in the figure are demagnified and the uncertainties from lens modeling have been included in the aforementioned results.

\begin{figure*}[htb]
    \centering
    \includegraphics[width = \linewidth]{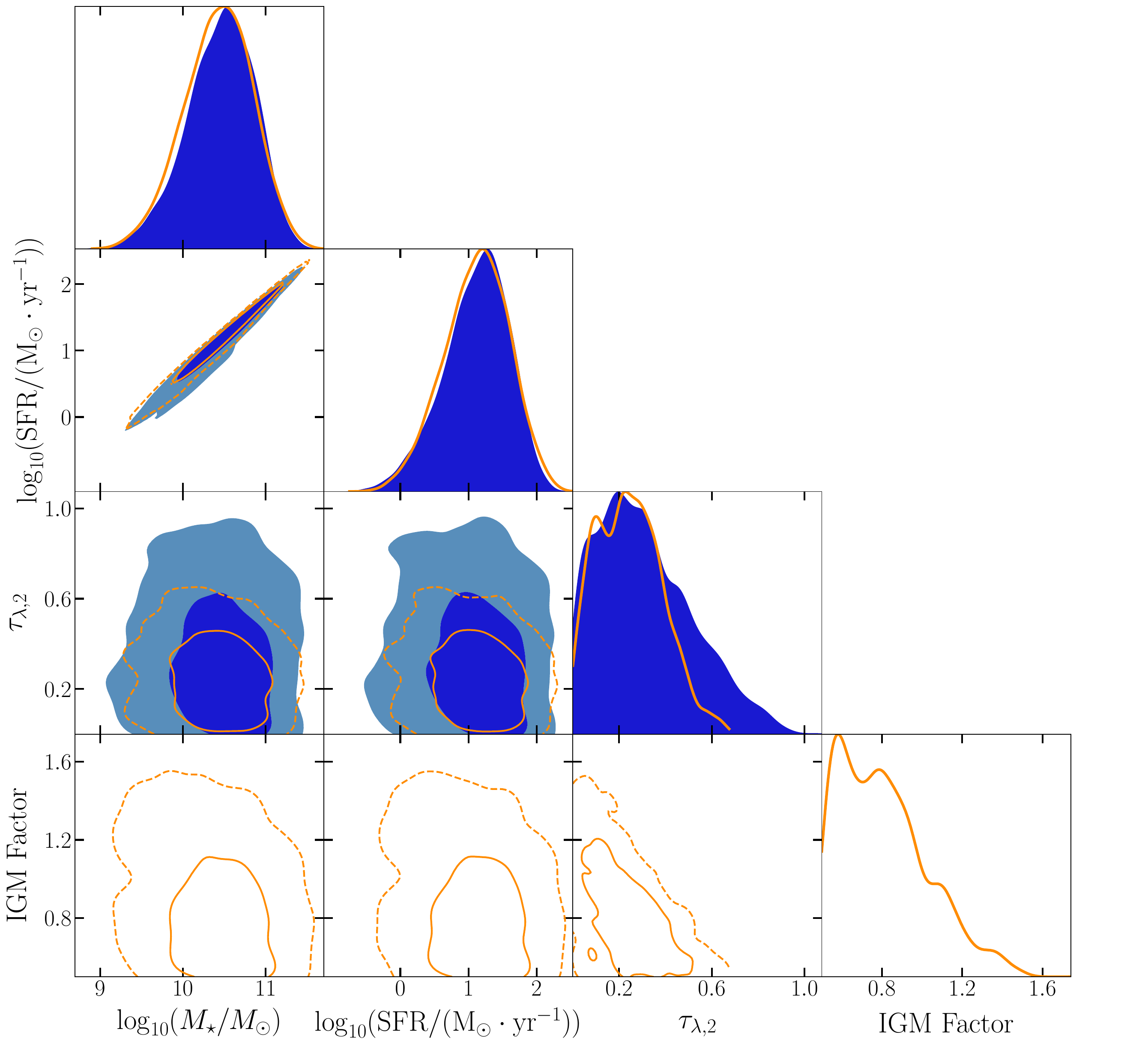}
    \caption{Comparison between SED models of CJ0145 in which the IGM factor is fixed as $1$ (blue filled contour) or a free parameter during fitting (orange open contour). The IGM absorption factor distribution peaks around $0.5$. Such a favoring of less attenuation does not impact the derived total stellar mass $M_{*}$ significantly.}
    \label{fig:corner2}
\end{figure*}

In addition, we experimented with a variant model with all $grz$ photometry, with an additional free parameter that tunes the intergalactic medium (IGM) absorption between the arc emission and the observer (based on the IGM absorption prescription by \citealt{Madau.etal.1995}); we chose a top-hat prior of $[0.5,2]$ for IGM absorption factor. All of our variant models favor a Landau attenuation factor that is less than expected. As shown in a selected example of CJ0145 in Figure \ref{fig:corner2}, the IGM absorption factor distribution has a 50th percentile value less than $1.0$. However, such a favoring of less attenuation does not impact the derived total stellar mass $M_{*}$ significantly. If we assume the posterior distributions of $M_{*}$ to be Gaussian, the derived $M_{*}$ from the variant models always falls in the $1 \sigma$ range of the fiducial model values.

\begin{table*}[htpb]
\caption{The Best-fit Values and Corresponding 68\% Confidence Intervals from SED Fitting of Demagnified Stellar Mass ($M_{\star}$), Demagnified Instantaneous SFR, and Diffuse Optical Depth ($\tau_{\lambda,2}$) for Galaxies in This Sample.}
    \centering
    \begin{tabular}{c|cccccc}
    \hline
    \hline
     & $\log_{10}(M_{\star}/M_{\odot} )$ & $\log_{10}(\rm SFR / (M_{\odot} \cdot yr^{-1} ))$  & $\tau_{\lambda, 2}$ \\
     \hline
     CJ0101  & $10.49^{+0.52}_{-0.53}$ & $1.21^{+0.57}_{-0.61}$ &  $0.82^{+0.19}_{-0.30}$ \\
      CJ0104 &  $10.75^{+0.40}_{-0.47}$ & $1.46^{+0.45}_{-0.57}$ & $0.61^{+0.22}_{-0.26}$ \\
      CJ0145 & $10.48^{+0.38}_{-0.44}$ & $1.16^{+0.43}_{-0.53}$ & $0.28^{+0.25}_{-0.19}$ \\
      CJ0516 &  $9.69^{+0.34}_{-0.40}$ & $0.39^{+0.40}_{-0.48}$ & $0.22^{+0.15}_{-0.14}$ \\
      CJ1356 &  $9.73^{+0.53}_{-0.58}$ & $0.43^{+0.61}_{-0.72}$ & $0.44^{+0.45}_{-0.31}$ \\
      \hline
    \end{tabular}
    \label{tab:sed_results}
\end{table*}

\subsection{Lens Modeling and Magnification}
\label{ssec:lens_model}

\begin{figure*}[t]
    \centering
    \includegraphics[width = \textwidth]{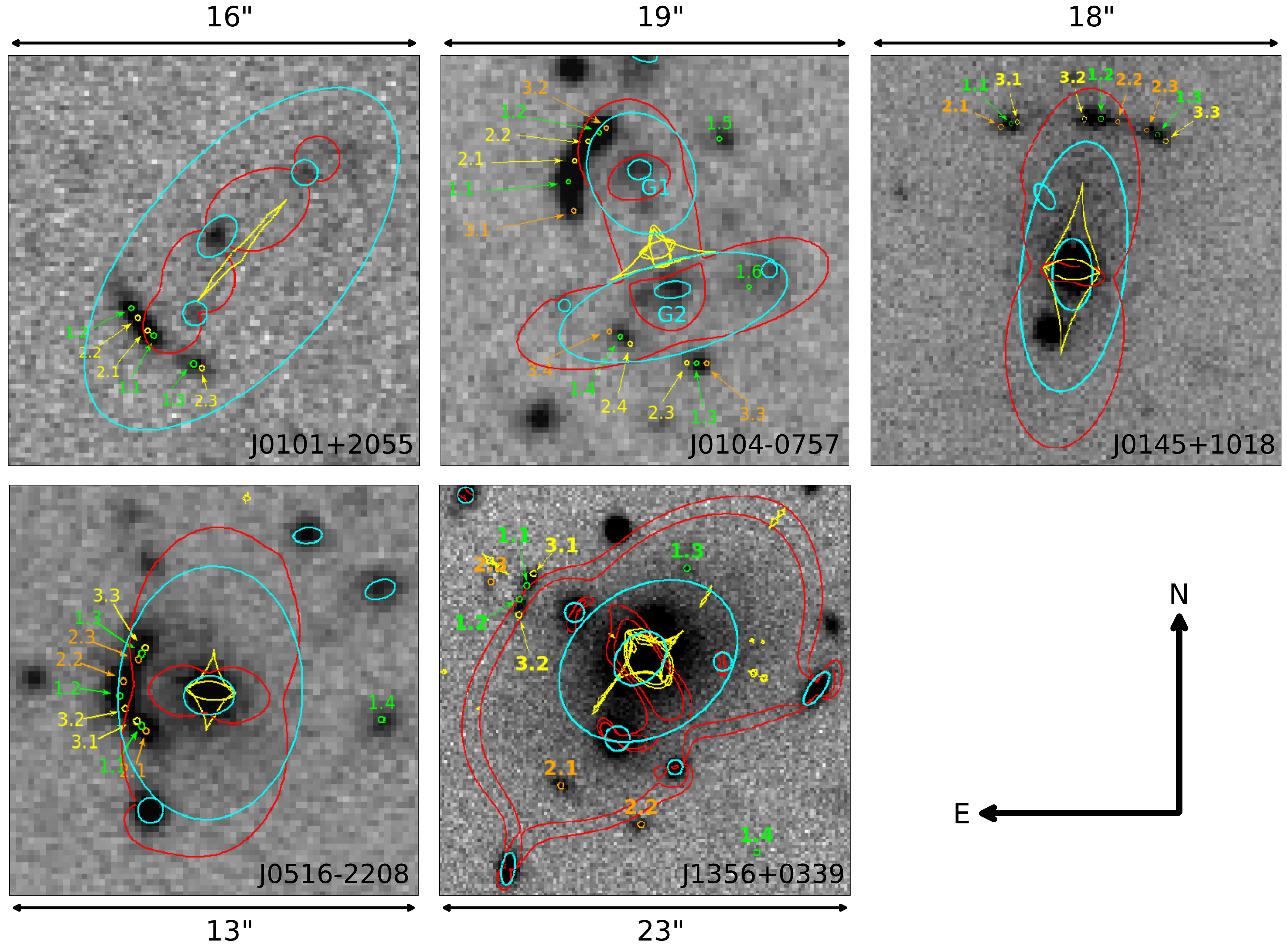}
    \caption{Lens models built with ground-based imaging. Different sets of positional constraints are denoted with different initial indexing and colors. The critical curve at the primary source redshift is in red (the additional inner critical curve for CJ1356 is for the second image family at lower redshift), and the caustic curve is in yellow. The sizes of the cyan ellipses show the relative scale of the halo masses. The field of view is chosen to best present each lensing system. Note that the second lensed image family in CJ1356 at $z = 2.17$ is marked in orange.  }
    \label{fig:lens_model}
\end{figure*}

To model the cluster-scale mass distribution that is responsible for the observed strong gravitational lensing, we used a parametric approach using \texttt{LENSTOOL}. In \texttt{LENSTOOL} (\citealt{Jullo.etal.2007.lenstool}), the mass distribution was modeled by dual pseudoisothermal ellipsoids (dPIEs; \citealt{Eliasdottir.etal.2007.pdie}) whose parameters and uncertainties were inferred through an MCMC approach. We closely followed the strong lens modeling process and model optimization strategy described in \citet{Sharon.etal.2020} and used the statistical uncertainties of our model following a $\chi$ statistic as in \citet{Mahler.etal.2020}. We selected the astrometric constraints for our modeling as multiple images of lensed features, identified from the color images. The best-fit model is defined as the one that minimizes the scatter between these observed image locations and their model-predicted counterparts in the image plane. In addition to the cluster-scale mass potentials, we fitted for galaxy-scale potentials whose normalization and radius parameters were determined by observed luminosity scaling relations \citep{Limousin.etal.2005}; cluster galaxies were selected from the DECaLS DR9 catalog as objects with extended morphologies whose colors are consistent with the expected red sequence at each lens redshift. The morphological parameters used for setting up galaxy-scale potentials, namely, position, ellipticity, and position angle were also taken from the DECALS DR9 catalog. 

In Figure \ref{fig:lens_model}, we overplot the positional constraints, individual fitted halos in the lens model, and critical curves on top of the imaging in the filter band in which the arc is the brightest. Each constraint family is labeled with the same color and first digit. The redshifts of the sources and lens are provided in Table \ref{tab:redshift}.

The lens models of CJ0101, CJ0145, and CJ0516 are the classic three-image arc configuration in which the image plane critical curve goes through the arc twice and a counterimage is potentially present. Note that although we do not have spectroscopic confirmations of the counterimages in the above three lensing systems, the best-fit model of CJ0516 requires the placement of a fourth constraint on the counterimage candidate. The lens models of CJ1356 and CJ0104 are more complicated and less obvious, and we provide a more detailed description of these systems below.

\subsubsection{CJ1356}
 In the lensing system of CJ1356, we find two lensed image families at two spectroscopically confirmed redshifts. The two images located in the northeast of the BCG (marked by constraint indices $1.1,1.2,3.1,3.2$) are spectroscopically confirmed to have a redshift at $z = 3.753$. In addition, two objects in the northwest (constraint $1.3$) and one object in the southwest (constraint $1.4$) are likely to be candidate lensed images from the same source as the $z = 3.753$ images, given their similar colors.
 
 We spectroscopically confirmed a second lensed image family at $z = 2.17$, which is composed of the images in the northeast (constraint $2.3$) and the image in the southeast (constraint $2.1$). There is a third image in the southeast (constraint $2.2$) for which we do not yet have spectroscopic confirmation. But given its color and geometrical location, we are confident that it belongs to the image family at $z = 2.17$.
 
 In the modeling process, we experimented with several astrometric constraint configurations with and without considering the candidate lensed images. The optimized lens model of this system requires the astrometric constraint placement shown in Figure \ref{fig:lens_model}. The model further suggests that only one of the two candidate sources in the northwest is likely from the green source family ($z = 3.753$).

\subsubsection{CJ0104}
In the lensing system of CJ0104, our best lens model suggests that the images are primarily lensed by a galaxy-scale or low-mass cluster-scale halo (shown as the north ellipse ``G1" in cyan in Figure \ref{fig:lens_model}) and a cluster-scale halo (shown as the south ellipse ``G2" in cyan in Figure \ref{fig:lens_model}). The galaxy hosted in the north galaxy-scale halo is spectroscopically confirmed at $z = 1.003$. The BCG of the cluster-scale halo is spectroscopically confirmed at $z = 0.858$.

The arc image (constraint $1.1,1.2,2.1,2.2,3.1,$and $ 3.2$) in the northeast and the image (constraint $1.3,2.3,3.3$) in the south are spectroscopically confirmed as belonging to the same source at $z = 3.480$. Given the similar color and location of the image (constraint $1.4,2.4,3.4$) in the southwest, it is likely to be part of the same image family. Our initial modeling suggests the placement of constraints $1.5$ and $1.6$. In the case of constraint $1.5$, we do find a candidate counterimage of similar color. In the case of constraint $1.6$, we suspect a counterimage hidden behind the foreground red cluster member, given the hint of green color in the RGB image. The model further suggests a candidate image located between the galaxy at $z = 1.003$ and the BCG at $z = 0.858$.

The precise modeling of the mass distribution in this system requires multiplane lens modeling, which is not yet implemented in \texttt{LENSTOOL}. Nevertheless, the purpose of lensing analysis in this study is to provide an estimation of the magnification factor for the lensed galaxy and assess the implication of projecting all the mass onto a single plane. We experimented with three instances in which all the mass potentials were uniformly placed at the lower, upper, and mean redshift ($z = 0.930$). Apart from the lens redshifts, we used the same initial inputs and priors for free parameters in the three models. At $z = 0.858$, the model returned a median magnification of $40_{-16}^{+55}$ for the bright arc (constraint $1.1,1.2,2.1,2.2,3.1,3.2$). At $z = 0.930$, the model returned a median magnification of $35_{-13}^{+54}$. At $z = 1.003$, the model returned a median magnification of $39_{-14}^{+47}$ (see Section \ref{sec:magnification} for the process of magnification estimation). The ranges of magnification factors estimated from the three model instances are statistically consistent with each other. We conclude that in the model of CJ0104, the single lens plane approximation does not affect the magnification estimate within the statistical uncertainties. We used the magnification estimated from the model at $z = 0.858$ for the analysis in this study. We show the lens model at $z = 0.858$ in Figure \ref{fig:lens_model}.

\subsubsection{Magnification}\label{sec:magnification}
To estimate the mean magnification, we calculated the ratio of the area in the image and source plane. For each confirmed arc image in each lensing system, we estimated the arc area in the image plane by taking the polygons that best describe the non-PSF-convolved GALFIT modeling image of the arc. We ray-traced its source plane counterpart using the deflection maps generated from the lens model, which describe the projection of the image plane positions onto the source plane. The source plane image size was estimated by taking the convex hull of pixel points deflected from the image plane. We obtained a mean total magnification for each of the lensed galaxies by summing the magnification of each image in the system. To estimate the uncertainties, we repeated the above procedures with 100 realizations of deflection maps that sample the lens model posterior probability distribution. The resulting magnifications and related uncertainties are shown in Table \ref{tab:redshift}. We applied this as a division factor to the results of the arc imaging analysis in the image plane, converting the derived physical parameter values from SED fitting to those in the source plane.

The magnification depends on the choice of the image plane polygon. With the limitations of imaging data from ground-based telescopes, we could not further refine the constraints in the lens model or the choice of polygons for magnification estimation. The reported magnifications and uncertainties are the best results obtained with the limited data available. For all but CJ1356 which is better constrained, the typical fractional uncertainties are $-50\%$ and $+100\%$.





\section{Discussion and Future Work}
\label{sec:discuss}



\begin{figure*}[h]
    \centering
    \includegraphics[width = \textwidth]{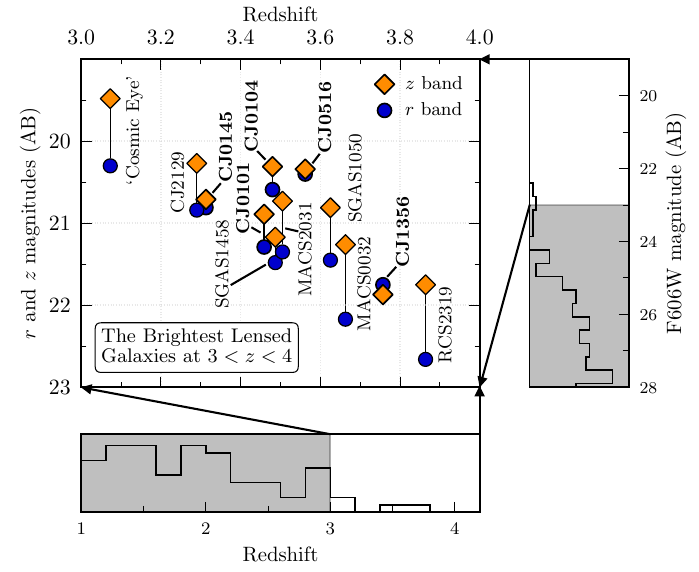}
    \caption{ \textit{Top left panel}: The total $r$- and $z$-band magnitudes for the brightest published lensed galaxies in the redshift interval $3 < z < 4$, and one further source (CJ2129) that will be presented in a future paper. These objects, including the newly discovered galaxies discussed in this paper (denoted in bold), represent the extreme tail in brightness for lensed sources in this redshift interval. Note that the photometry for CJ1356 is in the HSC $i$ and $y$ bands; however, we make an approximation of its $r$ and $z$-band magnitudes using the $i-y$ color for demonstration purposes here. \textit{Top right panel}: brightness distribution in the F606W band for $3 < z < 4$ sources from the MUSE spectroscopic sample of lensed sources behind 12 massive lensing clusters \citep{Richard.etal.2021}. \textit{Bottom panel}: redshift distribution of lensed sources in the CASSOWARY \citep{Stark.etal.2013.cassowary}, MEGASAURA \citep{Rigby.etal.2018.megasaura} and \citealt{Bayliss.etal.2011.26clus} samples. 
    }
    
    \label{fig:mags_objects3to4}
\end{figure*}

\begin{figure*}
    \centering
    \includegraphics[width = \textwidth]{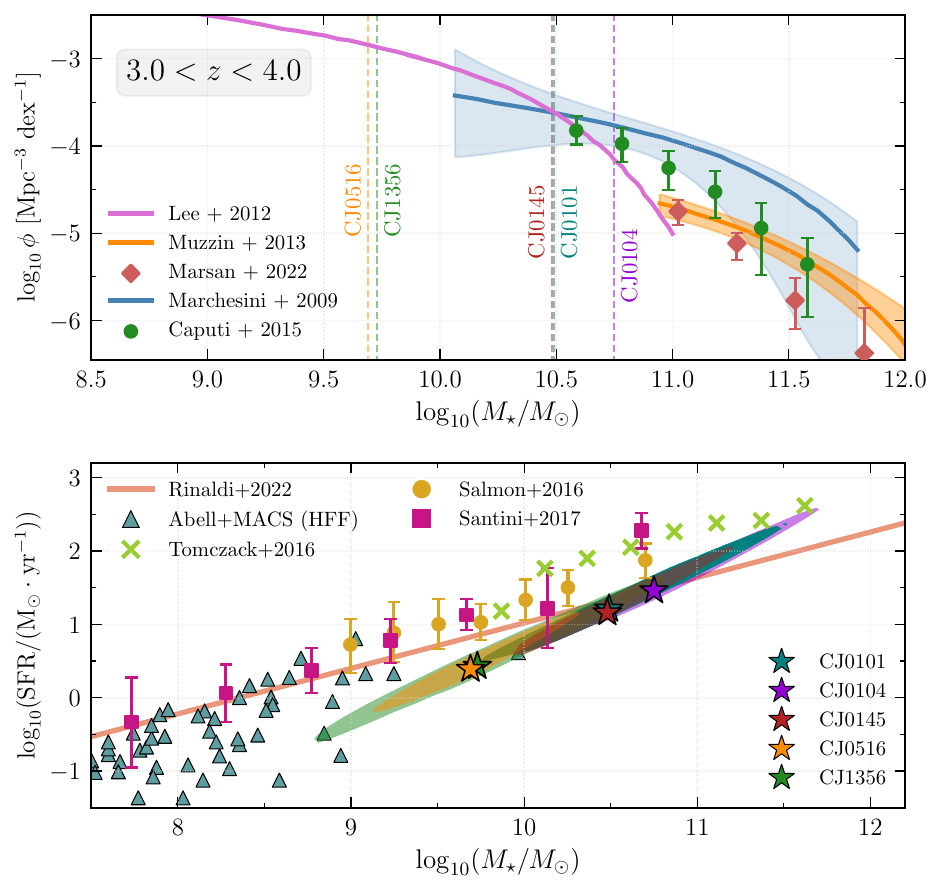}
    \caption{\textit{Top panel}: median estimates of rest-frame total stellar mass for galaxies in this sample (vertical dashed lines) and the star-forming galaxy SMF in the redshift interval $3 < z < 4$ from \citet{Lee.etal.2012}, \citet{Muzzin.etal.2013}, \citet{Marchesini.etal.2009}, \citet{Caputi.etal.2015}, and \citet{Marsan.etal.2022}. Shaded regions and error bars denote 68\% confidence intervals. \textit{Bottom panel}:
    The stellar mass $M_{\star}$ - SFR relation for galaxies in this sample and star-forming galaxies in the redshift interval $3 < z < 4$ \citep{Santini.etal.2017,inaldi.etal.2022,Salmon.etal.2015,Tomczak.etal.2016}. (Note that for Hubble Frontier Fields galaxies we use the $M_{\star}$ and SFR values derived in \citet{inaldi.etal.2022}.) The shaded regions show the 1$\sigma$ contours, while the stars show the median values. The galaxies in this sample lie along the main-sequence star-forming relation in this redshift interval.}
    \label{fig:ms_sfr_mf}
\end{figure*}



Figure~\ref{fig:mags_objects3to4} shows the $r$- and $z$-band magnitudes of the brightest published lensed galaxies found in the redshift interval $3 < z < 4$. The galaxies in bold denote those introduced in this paper.
To emphasize the lack of known bright lensed galaxies in this redshift interval, we show the distribution of redshifts and brightnesses of lensed galaxies in existing samples along the horizontal and vertical axes, respectively. The distribution along the horizontal axes shows the redshift distribution of lensed galaxies from the CASSOWARY \citep{Stark.etal.2013.cassowary}, MEGASAURA \citep{Rigby.etal.2018.megasaura} and \citet{Bayliss.etal.2011.26clus} samples. Furthermore, the distribution along the vertical axes shows the distribution of the F606W magnitudes for lensed galaxies in the redshift range $3 < z < 4$ from the MUSE spectroscopic sample of lensed sources \citep{Richard.etal.2021}. Note that the F606W band and $r/z$-band comparison is not precise but sufficient for illustration.


Another interesting feature in this lensed galaxy sample is the presence of Ly$\alpha$, or the lack thereof, as seen in the spectra (Figure~\ref{spectrums1} and \ref{spectrums2}). The Ly$\alpha$ emission is important for studying the physics of LyC escape in these galaxies. As the scattering of these photons is influenced by the neutral hydrogen reservoir in and around the galaxy, the shape of the Ly$\alpha$ emission line is altered and, in return, contains information on the neutral hydrogen. Hence, the Ly$\alpha$ emission line profile serves as an indirect indicator of the $f^{\rm Lyc}_{\rm esc}$, that is, the fraction of H I-ionizing LyC photons that escape from a galaxy \citep[e.g.,][]{Izotov.etal.2021, Gazagnes.etal.2020}. At redshifts of $z \gtrsim 4$, due to a large number of neutral hydrogen clumps, most LyC photons are absorbed \citep[e.g.,][]{Haardt.Madau.1996, Cowie.etal.1998, Vanzella.etal.2018}. However, by studying galaxies at $z \sim 3.0$, we are able to observe some of the ionizing photons. As a result, we can use these relatively lower-redshift galaxies as analogs of their epoch of reionization predecessors to study the escape of ionizing photons \citep[e.g.,][]{Rivera-Thorsen.etal.2017,Rivera.etal.2019.sunburst}. Understanding the cause of the variation in the presence of Ly$\alpha$, or the lack thereof, will require higher spatial resolution data, which will be enabled by space-based telescopes like the Hubble Space Telescope (HST) and JWST.


The relation between the stellar mass $M_{\star}$ and its instantaneous SFR has been proposed to be a fundamental relation of galaxies containing information on the evolutionary state of the galaxy and variations in the star formation histories \citep[e.g.,][]{Matthee.etal.2019}. This relationship has been observed over high orders of magnitude in stellar mass \citep{Santini.etal.2009} and from redshifts $z=0$ to $6$ \citep[e.g.,][]{Tasca.etal.2015}. 

The bottom panel of Figure~\ref{fig:ms_sfr_mf} shows the stellar mass - SFR relation for main-sequence galaxies in the redshift interval $3 < z < 4$. The figure shows a compilation of results from numerous works and this lensed galaxy sample for comparison. Galaxies in this sample appear to be fairly typical, as they lie along the main-sequence star-forming relation. Furthermore, to contextualize these galaxies within the stellar mass function (SMF) of star-forming galaxies in this redshift interval, the top panel of Figure~\ref{fig:ms_sfr_mf} shows various estimates of the SMF and the stellar masses of our galaxy sample as vertical dashed lines. The deduced stellar masses are comparable to the characteristic galaxy mass at these redshifts \citep[e.g.,][]{Lee.etal.2012}. With more detailed photometry in future work, we will have better constraints on the star formation history of these galaxies and thus reduce the uncertainty in our stellar mass and SFR estimates.



These strongly lensed galaxies are some of the brightest galaxies found in the redshift interval $3 < z < 4$ (see Figure~\ref{fig:mags_objects3to4}). Due to their high surface brightness (and hence high S/Ns), even with limited ground-based photometry, we are able to estimate some basic properties of these galaxies and their foreground lensing systems. Note that galaxy properties like SFR, stellar mass, and physical size scale directly with the lensing magnification. Thus, having a robust lens model is important. Space-based telescopes HST and JWST have high spatial resolution and thus enable the construction of robust lens models. Recent studies of strong lensing systems using HST and JWST \citep[e.g.,][]{Rivera.etal.2019.sunburst,Wu.etal.2022.jwstlens,Mahler.etal.2022.jwstsmacs, Claeyssens.etal.2022.jwst} were able to discern morphology and star formation at the smallest scales. Furthermore, in this redshift interval, all the rest-frame optical nebular emission lines fall in the JWST NIRSpec bandpass. Also, the UV emission and absorption lines, which probe the massive stellar populations and outflowing winds, are in the bandpass of ground-based optical integral field units like MUSE (Very Large Telescope) and KCWI (W. M. Keck Observatory). Thus, the lensed galaxies in this sample are excellent future targets for such follow-up observations.

\section{Summary}
\label{sec:summary}

We report the discoveries of five bright, strongly-lensed galaxies at $3 < z < 4$: COOL J0101+2055 ($z_{source} = 3.459$; $z_{lens} = 0.871$), COOL J0104-0757 ($z_{source} = 3.480$; $z_{lens1} = 1.004$; $z_{lens2} = 0.858$), COOL J0145+1018 ($z_{source} = 3.310$; $z_{lens} = 0.463$), COOL J0516-2208 ($z_{source} = 3.549$; $z_{lens} = 0.658$), and COOL J1356+0339 ($z_{source} = 3.753$; $z_{lens} = 0.262$). These galaxies have {\it r}- and {\it z}-band AB magnitudes brighter than $21.81$, doubling the number of bright lensed galaxies currently known in this redshift interval.  We characterized the lensed galaxies using ground-based {\it grz}/{\it giy} imaging and optical spectroscopy. Using \texttt{Prospector}, we performed stellar population synthesis modeling with model-based photometry magnitudes and derived stellar masses, dust content, and SFRs. Using \texttt{LENSTOOL}, we performed dPIE halo-based lens mass modeling with ground-based imaging, which implies median source magnifications in our sample of $\sim 29 - 180$. Combining these analyses, we derived the sample galaxies' stellar masses in the range $\rm log_{10}(M_{*}/M_{\odot}) \sim 9.69 - 10.75$ and SFRs in the youngest age bin in the range $\rm log_{10}(SFR/(M_{\odot}\cdot yr^{-1})) \sim 0.39 - 1.46$, placing the sample galaxies on the massive end of the star-forming main sequence. In summary, the five galaxies presented in this sample are rare in terms of their brightness and common in terms of their stellar mass in the redshift interval $3 < z < 4$. In addition, three galaxies in our sample have strong Ly$\alpha$ emission, enabling unique opportunities to study Ly$\alpha$ emitters at high redshifts.

\section*{Acknowledgements}
We thank Jason J. Lin, Elisabeth Medina and Daniel J. Kavin Stein for their contribution to the initial lens searches.

This research has made use of the NASA/IPAC Extragalactic Database, which is funded by the National Aeronautics and Space Administration and operated by the California Institute of Technology.

This paper is based on data gathered with the 6.5 m Magellan Telescopes located at Las Campanas Observatory, Chile. Magellan observing time for this program was granted by the time allocation committees of the University of Chicago and the University of Michigan.

Based on observations made with the Nordic Optical Telescope, owned in collaboration by the University of Turku and Aarhus University, and operated jointly by Aarhus University, the University of Turku and the University of Oslo, representing Denmark, Finland and Norway, the University of Iceland and Stockholm University at the Observatorio del Roque de los Muchachos, La Palma, Spain, of the Instituto de Astrofisica de Canarias. The data presented here were obtained in part with ALFOSC, which is provided by the Instituto de Astrofisica de Andalucia (IAA) under a joint agreement with the University of Copenhagen and NOT.

This work is supported by The College Undergraduate program at the University of Chicago, and the Department of Astronomy and Astrophysics at the University of Chicago.

The Legacy Surveys consist of three individual and complementary projects: the Dark Energy Camera Legacy Survey (DECaLS; Proposal ID 2014B-0404; PIs: David Schlegel and Arjun Dey), the Beijing-Arizona Sky Survey (BASS; NOAO Prop. ID 2015A-0801; PIs: Zhou Xu and Xiaohui Fan), and the Mayall z-band Legacy Survey (MzLS; Prop. ID 2016A-0453; PI: Arjun Dey). DECaLS, BASS and MzLS together include data obtained, respectively, at the Blanco telescope, Cerro Tololo Inter-American Observatory, NSF’s NOIRLab; the Bok telescope, Steward Observatory, University of Arizona; and the Mayall telescope, Kitt Peak National Observatory, NOIRLab. The Legacy Surveys project is honored to be permitted to conduct astronomical research on Iolkam Du’ag (Kitt Peak), a mountain with particular significance to the Tohono O’odham Nation.

NOIRLab is operated by the Association of Universities for Research in Astronomy (AURA) under a cooperative agreement with the National Science Foundation.

This project used data obtained with the Dark Energy Camera (DECam), which was constructed by the Dark Energy Survey (DES) collaboration. Funding for the DES Projects has been provided by the U.S. Department of Energy, the U.S. National Science Foundation, the Ministry of Science and Education of Spain, the Science and Technology Facilities Council of the United Kingdom, the Higher Education Funding Council for England, the National Center for Supercomputing Applications at the University of Illinois at Urbana-Champaign, the Kavli Institute of Cosmological Physics at the University of Chicago, Center for Cosmology and Astro-Particle Physics at the Ohio State University, the Mitchell Institute for Fundamental Physics and Astronomy at Texas A\&M University, Financiadora de Estudos e Projetos, Fundacao Carlos Chagas Filho de Amparo, Financiadora de Estudos e Projetos, Fundacao Carlos Chagas Filho de Amparo a Pesquisa do Estado do Rio de Janeiro, Conselho Nacional de Desenvolvimento Cientifico e Tecnologico and the Ministerio da Ciencia, Tecnologia e Inovacao, the Deutsche Forschungsgemeinschaft and the Collaborating Institutions in the Dark Energy Survey. The Collaborating Institutions are Argonne National Laboratory, the University of California at Santa Cruz, the University of Cambridge, Centro de Investigaciones Energeticas, Medioambientales y Tecnologicas-Madrid, the University of Chicago, University College London, the DES-Brazil Consortium, the University of Edinburgh, the Eidgenossische Technische Hochschule (ETH) Zurich, Fermi National Accelerator Laboratory, the University of Illinois at Urbana-Champaign, the Institut de Ciencies de l’Espai (IEEC/CSIC), the Institut de Fisica d’Altes Energies, Lawrence Berkeley National Laboratory, the Ludwig Maximilians Universitat Munchen and the associated Excellence Cluster Universe, the University of Michigan, NSF’s NOIRLab, the University of Nottingham, the Ohio State University, the University of Pennsylvania, the University of Portsmouth, SLAC National Accelerator Laboratory, Stanford University, the University of Sussex, and Texas A\&M University.

The Legacy Surveys imaging of the DESI footprint is supported by the Director, Office of Science, Office of High Energy Physics of the U.S. Department of Energy under Contract No. DE-AC02- 05CH1123, by the National Energy Research Scientific Computing Center, a DOE Office of Science User Facility under the same contract; and by the U.S. National Science Foundation, Division of Astronomical Sciences under Contract No. AST-0950945 to NOIRLab.

\facilities{Magellan Telescopes 6.5m (Clay/LDSS3C, Baade/IMACS, Baade/FIRE), Nordic Optical Telescope 2.56m/ALFOSC}

\software{\texttt{NumPy}\citep{numpy_paper}, 
          \texttt{matplotlib} \citep{matplotlib_paper},
          \texttt{astropy} \citep{Astropy}, 
          \texttt{scipy} \citep{2020SciPy-NMeth},
          \texttt{SEDpy} \citep{SEDpy},
          \texttt{Prospector} \citep{Johnson.etal.2021.Prospector}, 
          \texttt{SExtractor} \citep{sextractor},
          \texttt{Lenstool} \citep{Jullo.etal.2007.lenstool}, 
          \texttt{GALFIT} \citep{Peng.etal.2010}
          }



\bibliography{main}

\begin{thebibliography}{}
\expandafter\ifx\csname natexlab\endcsname\relax\def\natexlab#1{#1}\fi
\providecommand{\url}[1]{\href{#1}{#1}}
\providecommand{\dodoi}[1]{doi:~\href{http://doi.org/#1}{\nolinkurl{#1}}}
\providecommand{\doeprint}[1]{\href{http://ascl.net/#1}{\nolinkurl{http://ascl.net/#1}}}
\providecommand{\doarXiv}[1]{\href{https://arxiv.org/abs/#1}{\nolinkurl{https://arxiv.org/abs/#1}}}

\bibitem[{{Ahumada} {et~al.}(2020){Ahumada}, {Prieto}, {Almeida}, {Anders},
  {Anderson}, {Andrews}, {Anguiano}, {Arcodia}, {Armengaud}, {Aubert}, {Avila},
  {Avila-Reese}, {Badenes}, {Balland}, {Barger}, {Barrera-Ballesteros}, {Basu},
  {Bautista}, {Beaton}, {Beers}, {Benavides}, {Bender}, {Bernardi}, {Bershady},
  {Beutler}, {Bidin}, {Bird}, {Bizyaev}, {Blanc}, {Blanton}, {Boquien},
  {Borissova}, {Bovy}, {Brandt}, {Brinkmann}, {Brownstein}, {Bundy}, {Bureau},
  {Burgasser}, {Burtin}, {Cano-D{\'\i}az}, {Capasso}, {Cappellari}, {Carrera},
  {Chabanier}, {Chaplin}, {Chapman}, {Cherinka}, {Chiappini}, {Doohyun Choi},
  {Chojnowski}, {Chung}, {Clerc}, {Coffey}, {Comerford}, {Comparat}, {da
  Costa}, {Cousinou}, {Covey}, {Crane}, {Cunha}, {Ilha}, {Dai}, {Damsted},
  {Darling}, {Davidson}, {Davies}, {Dawson}, {De}, {de la Macorra}, {De Lee},
  {Queiroz}, {Deconto Machado}, {de la Torre}, {Dell'Agli}, {du Mas des
  Bourboux}, {Diamond-Stanic}, {Dillon}, {Donor}, {Drory}, {Duckworth},
  {Dwelly}, {Ebelke}, {Eftekharzadeh}, {Davis Eigenbrot}, {Elsworth},
  {Eracleous}, {Erfanianfar}, {Escoffier}, {Fan}, {Farr},
  {Fern{\'a}ndez-Trincado}, {Feuillet}, {Finoguenov}, {Fofie},
  {Fraser-McKelvie}, {Frinchaboy}, {Fromenteau}, {Fu}, {Galbany}, {Garcia},
  {Garc{\'\i}a-Hern{\'a}ndez}, {Oehmichen}, {Ge}, {Maia}, {Geisler}, {Gelfand},
  {Goddy}, {Gonzalez-Perez}, {Grabowski}, {Green}, {Grier}, {Guo}, {Guy},
  {Harding}, {Hasselquist}, {Hawken}, {Hayes}, {Hearty}, {Hekker}, {Hogg},
  {Holtzman}, {Horta}, {Hou}, {Hsieh}, {Huber}, {Hunt}, {Chitham}, {Imig},
  {Jaber}, {Angel}, {Johnson}, {Jones}, {J{\"o}nsson}, {Jullo}, {Kim},
  {Kinemuchi}, {Kirkpatrick}, {Kite}, {Klaene}, {Kneib}, {Kollmeier}, {Kong},
  {Kounkel}, {Krishnarao}, {Lacerna}, {Lan}, {Lane}, {Law}, {Le Goff}, {Leung},
  {Lewis}, {Li}, {Lian}, {Lin}, {Long}, {Longa-Pe{\~n}a}, {Lundgren}, {Lyke},
  {Ted Mackereth}, {MacLeod}, {Majewski}, {Manchado}, {Maraston}, {Martini},
  {Masseron}, {Masters}, {Mathur}, {McDermid}, {Merloni}, {Merrifield},
  {M{\'e}sz{\'a}ros}, {Miglio}, {Minniti}, {Minsley}, {Miyaji}, {Mohammad},
  {Mosser}, {Mueller}, {Muna}, {Mu{\~n}oz-Guti{\'e}rrez}, {Myers}, {Nadathur},
  {Nair}, {Nandra}, {do Nascimento}, {Nevin}, {Newman}, {Nidever}, {Nitschelm},
  {Noterdaeme}, {O'Connell}, {Olmstead}, {Oravetz}, {Oravetz}, {Osorio},
  {Pace}, {Padilla}, {Palanque-Delabrouille}, {Palicio}, {Pan}, {Pan},
  {Parker}, {Paviot}, {Peirani}, {Ram{\'r}ez}, {Penny}, {Percival},
  {Perez-Fournon}, {P{\'e}rez-R{\`a}fols}, {Petitjean}, {Pieri},
  {Pinsonneault}, {Poovelil}, {Povick}, {Prakash}, {Price-Whelan}, {Raddick},
  {Raichoor}, {Ray}, {Rembold}, {Rezaie}, {Riffel}, {Riffel}, {Rix}, {Robin},
  {Roman-Lopes}, {Rom{\'a}n-Z{\'u}{\~n}iga}, {Rose}, {Ross}, {Rossi},
  {Rowlands}, {Rubin}, {Salvato}, {S{\'a}nchez}, {S{\'a}nchez-Menguiano},
  {S{\'a}nchez-Gallego}, {Sayres}, {Schaefer}, {Schiavon}, {Schimoia},
  {Schlafly}, {Schlegel}, {Schneider}, {Schultheis}, {Schwope}, {Seo},
  {Serenelli}, {Shafieloo}, {Shamsi}, {Shao}, {Shen}, {Shetrone}, {Shirley},
  {Aguirre}, {Simon}, {Skrutskie}, {Slosar}, {Smethurst}, {Sobeck}, {Sodi},
  {Souto}, {Stark}, {Stassun}, {Steinmetz}, {Stello}, {Stermer},
  {Storchi-Bergmann}, {Streblyanska}, {Stringfellow}, {Stutz}, {Su{\'a}rez},
  {Sun}, {Taghizadeh-Popp}, {Talbot}, {Tayar}, {Thakar}, {Theriault}, {Thomas},
  {Thomas}, {Tinker}, {Tojeiro}, {Toledo}, {Tremonti}, {Troup}, {Tuttle},
  {Unda-Sanzana}, {Valentini}, {Vargas-Gonz{\'a}lez}, {Vargas-Maga{\~n}a},
  {V{\'a}zquez-Mata}, {Vivek}, {Wake}, {Wang}, {Weaver}, {Weijmans}, {Wild},
  {Wilson}, {Wilson}, {Wolthuis}, {Wood-Vasey}, {Yan}, {Yang}, {Y{\`e}che},
  {Zamora}, {Zarrouk}, {Zasowski}, {Zhang}, {Zhao}, {Zhao}, {Zheng}, {Zheng},
  {Zhu}, \& {Zou}}]{SDSS_dr16}
{Ahumada}, R., {Prieto}, C.~A., {Almeida}, A., {et~al.} 2020, \apjs, 249, 3,
  \dodoi{10.3847/1538-4365/ab929e}

\bibitem[{{Aihara} {et~al.}(2021){Aihara}, {AlSayyad}, {Ando}, {Armstrong},
  {Bosch}, {Egami}, {Furusawa}, {Furusawa}, {Harasawa}, {Harikane}, {Hsieh},
  {Ikeda}, {Ito}, {Iwata}, {Kodama}, {Koike}, {Kokubo}, {Komiyama}, {Li},
  {Liang}, {Lin}, {Lupton}, {Lust}, {MacArthur}, {Mawatari}, {Mineo},
  {Miyatake}, {Miyazaki}, {More}, {Morishima}, {Murayama}, {Nakajima},
  {Nakata}, {Nishizawa}, {Oguri}, {Okabe}, {Okura}, {Ono}, {Osato}, {Ouchi},
  {Pan}, {Plazas Malag{\'o}n}, {Price}, {Reed}, {Rykoff}, {Shibuya},
  {Simunovic}, {Strauss}, {Sugimori}, {Suto}, {Suzuki}, {Takada}, {Takagi},
  {Takata}, {Takita}, {Tanaka}, {Tang}, {Taranu}, {Terai}, {Toba}, {Turner},
  {Uchiyama}, {Vijarnwannaluk}, {Waters}, {Yamada}, {Yamamoto}, \&
  {Yamashita}}]{Aihara.etal.2021.DR3}
{Aihara}, H., {AlSayyad}, Y., {Ando}, M., {et~al.} 2021, arXiv e-prints,
  arXiv:2108.13045.
\newblock \doarXiv{2108.13045}

\bibitem[{{Alavi} {et~al.}(2016){Alavi}, {Siana}, {Richard}, {Rafelski},
  {Jauzac}, {Limousin}, {Freeman}, {Scarlata}, {Robertson}, {Stark}, {Teplitz},
  \& {Desai}}]{alavi2016}
{Alavi}, A., {Siana}, B., {Richard}, J., {et~al.} 2016, \apj, 832, 56,
  \dodoi{10.3847/0004-637X/832/1/56}

\bibitem[{{Astropy Collaboration} {et~al.}(2022){Astropy Collaboration},
  {Price-Whelan}, {Lim}, {Earl}, {Starkman}, {Bradley}, {Shupe}, {Patil},
  {Corrales}, {Brasseur}, {N{\"o}the}, {Donath}, {Tollerud}, {Morris},
  {Ginsburg}, {Vaher}, {Weaver}, {Tocknell}, {Jamieson}, {van Kerkwijk},
  {Robitaille}, {Merry}, {Bachetti}, {G{\"u}nther}, {Aldcroft},
  {Alvarado-Montes}, {Archibald}, {B{\'o}di}, {Bapat}, {Barentsen},
  {Baz{\'a}n}, {Biswas}, {Boquien}, {Burke}, {Cara}, {Cara}, {Conroy},
  {Conseil}, {Craig}, {Cross}, {Cruz}, {D'Eugenio}, {Dencheva}, {Devillepoix},
  {Dietrich}, {Eigenbrot}, {Erben}, {Ferreira}, {Foreman-Mackey}, {Fox},
  {Freij}, {Garg}, {Geda}, {Glattly}, {Gondhalekar}, {Gordon}, {Grant},
  {Greenfield}, {Groener}, {Guest}, {Gurovich}, {Handberg}, {Hart},
  {Hatfield-Dodds}, {Homeier}, {Hosseinzadeh}, {Jenness}, {Jones}, {Joseph},
  {Kalmbach}, {Karamehmetoglu}, {Ka{\l}uszy{\'n}ski}, {Kelley}, {Kern},
  {Kerzendorf}, {Koch}, {Kulumani}, {Lee}, {Ly}, {Ma}, {MacBride}, {Maljaars},
  {Muna}, {Murphy}, {Norman}, {O'Steen}, {Oman}, {Pacifici}, {Pascual},
  {Pascual-Granado}, {Patil}, {Perren}, {Pickering}, {Rastogi}, {Roulston},
  {Ryan}, {Rykoff}, {Sabater}, {Sakurikar}, {Salgado}, {Sanghi}, {Saunders},
  {Savchenko}, {Schwardt}, {Seifert-Eckert}, {Shih}, {Jain}, {Shukla}, {Sick},
  {Simpson}, {Singanamalla}, {Singer}, {Singhal}, {Sinha}, {Sip{\H{o}}cz},
  {Spitler}, {Stansby}, {Streicher}, {{\v{S}}umak}, {Swinbank}, {Taranu},
  {Tewary}, {Tremblay}, {Val-Borro}, {Van Kooten}, {Vasovi{\'c}}, {Verma}, {de
  Miranda Cardoso}, {Williams}, {Wilson}, {Winkel}, {Wood-Vasey}, {Xue},
  {Yoachim}, {Zhang}, {Zonca}, \& {Astropy Project Contributors}}]{Astropy}
{Astropy Collaboration}, {Price-Whelan}, A.~M., {Lim}, P.~L., {et~al.} 2022,
  \apj, 935, 167, \dodoi{10.3847/1538-4357/ac7c74}

\bibitem[{{Atek} {et~al.}(2008){Atek}, {Kunth}, {Hayes}, {{\"O}stlin}, \&
  {Mas-Hesse}}]{atek2008}
{Atek}, H., {Kunth}, D., {Hayes}, M., {{\"O}stlin}, G., \& {Mas-Hesse}, J.~M.
  2008, \aap, 488, 491, \dodoi{10.1051/0004-6361:200809527}

\bibitem[{{Bayliss}(2012)}]{matt2012}
{Bayliss}, M.~B. 2012, \apj, 744, 156, \dodoi{10.1088/0004-637X/744/2/156}

\bibitem[{{Bayliss} {et~al.}(2011{\natexlab{a}}){Bayliss}, {Gladders}, {Oguri},
  {Hennawi}, {Sharon}, {Koester}, \& {Dahle}}]{matt2011}
{Bayliss}, M.~B., {Gladders}, M.~D., {Oguri}, M., {et~al.} 2011{\natexlab{a}},
  \apjl, 727, L26, \dodoi{10.1088/2041-8205/727/1/L26}

\bibitem[{{Bayliss} {et~al.}(2011{\natexlab{b}}){Bayliss}, {Hennawi},
  {Gladders}, {Koester}, {Sharon}, {Dahle}, \&
  {Oguri}}]{Bayliss.etal.2011.26clus}
{Bayliss}, M.~B., {Hennawi}, J.~F., {Gladders}, M.~D., {et~al.}
  2011{\natexlab{b}}, \apjs, 193, 8, \dodoi{10.1088/0067-0049/193/1/8}

\bibitem[{{Bayliss} {et~al.}(2014){Bayliss}, {Rigby}, {Sharon}, {Wuyts},
  {Florian}, {Gladders}, {Johnson}, \& {Oguri}}]{Bayliss.etal.2014.sgas1050}
{Bayliss}, M.~B., {Rigby}, J.~R., {Sharon}, K., {et~al.} 2014, \apj, 790, 144,
  \dodoi{10.1088/0004-637X/790/2/144}

\bibitem[{{Bertin} \& {Arnouts}(1996)}]{sextractor}
{Bertin}, E., \& {Arnouts}, S. 1996, \aaps, 117, 393,
  \dodoi{10.1051/aas:1996164}

\bibitem[{{Byler} {et~al.}(2020){Byler}, {Kewley}, {Rigby}, {Acharyya}, {Berg},
  {Bayliss}, \& {Sharon}}]{byler20}
{Byler}, N., {Kewley}, L.~J., {Rigby}, J.~R., {et~al.} 2020, \apj, 893, 1,
  \dodoi{10.3847/1538-4357/ab7ea9}

\bibitem[{{Calzetti} {et~al.}(2000){Calzetti}, {Armus}, {Bohlin}, {Kinney},
  {Koornneef}, \& {Storchi-Bergmann}}]{Calzetti.etal.2000}
{Calzetti}, D., {Armus}, L., {Bohlin}, R.~C., {et~al.} 2000, \apj, 533, 682,
  \dodoi{10.1086/308692}

\bibitem[{{Calzetti} {et~al.}(1994){Calzetti}, {Kinney}, \&
  {Storchi-Bergmann}}]{calzetti1994}
{Calzetti}, D., {Kinney}, A.~L., \& {Storchi-Bergmann}, T. 1994, \apj, 429,
  582, \dodoi{10.1086/174346}

\bibitem[{{Caputi} {et~al.}(2015){Caputi}, {Ilbert}, {Laigle}, {McCracken}, {Le
  F{\`e}vre}, {Fynbo}, {Milvang-Jensen}, {Capak}, {Salvato}, \&
  {Taniguchi}}]{Caputi.etal.2015}
{Caputi}, K.~I., {Ilbert}, O., {Laigle}, C., {et~al.} 2015, \apj, 810, 73,
  \dodoi{10.1088/0004-637X/810/1/73}

\bibitem[{{Christensen} {et~al.}(2012){Christensen}, {Richard}, {Hjorth},
  {Milvang-Jensen}, {Laursen}, {Limousin}, {Dessauges-Zavadsky}, {Grillo}, \&
  {Ebeling}}]{Christensen12}
{Christensen}, L., {Richard}, J., {Hjorth}, J., {et~al.} 2012, \mnras, 427,
  1953, \dodoi{10.1111/j.1365-2966.2012.22006.x}

\bibitem[{{Claeyssens} {et~al.}(2022){Claeyssens}, {Adamo}, {Richard},
  {Mahler}, {Messa}, \& {Dessauges-Zavadsky}}]{Claeyssens.etal.2022.jwst}
{Claeyssens}, A., {Adamo}, A., {Richard}, J., {et~al.} 2022, arXiv e-prints,
  arXiv:2208.10450.
\newblock \doarXiv{2208.10450}

\bibitem[{{Conroy}(2013)}]{Conroy.etal.2013}
{Conroy}, C. 2013, \araa, 51, 393, \dodoi{10.1146/annurev-astro-082812-141017}

\bibitem[{{Conroy} \& {Gunn}(2010)}]{Conroy.etal.2010}
{Conroy}, C., \& {Gunn}, J.~E. 2010, \apj, 712, 833,
  \dodoi{10.1088/0004-637X/712/2/833}

\bibitem[{{Conroy} {et~al.}(2009){Conroy}, {Gunn}, \&
  {White}}]{Conroy.etal.2009}
{Conroy}, C., {Gunn}, J.~E., \& {White}, M. 2009, \apj, 699, 486,
  \dodoi{10.1088/0004-637X/699/1/486}

\bibitem[{{Coppin} {et~al.}(2007){Coppin}, {Swinbank}, {Neri}, {Cox}, {Smail},
  {Ellis}, {Geach}, {Siana}, {Teplitz}, {Dye}, {Kneib}, {Edge}, \&
  {Richard}}]{Coppin07}
{Coppin}, K.~E.~K., {Swinbank}, A.~M., {Neri}, R., {et~al.} 2007, \apj, 665,
  936, \dodoi{10.1086/519789}

\bibitem[{{Cowie} \& {Hu}(1998)}]{Cowie.etal.1998}
{Cowie}, L.~L., \& {Hu}, E.~M. 1998, \aj, 115, 1319, \dodoi{10.1086/300309}

\bibitem[{{Dahle} {et~al.}(2016){Dahle}, {Aghanim}, {Guennou}, {Hudelot},
  {Kneissl}, {Pointecouteau}, {Beelen}, {Bayliss}, {Douspis}, {Nesvadba},
  {Hempel}, {Gronke}, {Burenin}, {Dole}, {Harrison}, {Mazzotta}, \&
  {Sunyaev}}]{Dahle.etal.2016}
{Dahle}, H., {Aghanim}, N., {Guennou}, L., {et~al.} 2016, \aap, 590, L4,
  \dodoi{10.1051/0004-6361/201628297}

\bibitem[{{Dawson} {et~al.}(2016){Dawson}, {Kneib}, {Percival}, {Alam},
  {Albareti}, {Anderson}, {Armengaud}, {Aubourg}, {Bailey}, {Bautista},
  {Berlind}, {Bershady}, {Beutler}, {Bizyaev}, {Blanton}, {Blomqvist},
  {Bolton}, {Bovy}, {Brandt}, {Brinkmann}, {Brownstein}, {Burtin}, {Busca},
  {Cai}, {Chuang}, {Clerc}, {Comparat}, {Cope}, {Croft}, {Cruz-Gonzalez}, {da
  Costa}, {Cousinou}, {Darling}, {de la Macorra}, {de la Torre}, {Delubac}, {du
  Mas des Bourboux}, {Dwelly}, {Ealet}, {Eisenstein}, {Eracleous}, {Escoffier},
  {Fan}, {Finoguenov}, {Font-Ribera}, {Frinchaboy}, {Gaulme}, {Georgakakis},
  {Green}, {Guo}, {Guy}, {Ho}, {Holder}, {Huehnerhoff}, {Hutchinson}, {Jing},
  {Jullo}, {Kamble}, {Kinemuchi}, {Kirkby}, {Kitaura}, {Klaene}, {Laher},
  {Lang}, {Laurent}, {Le Goff}, {Li}, {Liang}, {Lima}, {Lin}, {Lin}, {Lin},
  {Long}, {Lundgren}, {MacDonald}, {Geimba Maia}, {Malanushenko},
  {Malanushenko}, {Mariappan}, {McBride}, {McGreer}, {M{\'e}nard}, {Merloni},
  {Meza}, {Montero-Dorta}, {Muna}, {Myers}, {Nandra}, {Naugle}, {Newman},
  {Noterdaeme}, {Nugent}, {Ogando}, {Olmstead}, {Oravetz}, {Oravetz},
  {Padmanabhan}, {Palanque-Delabrouille}, {Pan}, {Parejko}, {P{\^a}ris},
  {Peacock}, {Petitjean}, {Pieri}, {Pisani}, {Prada}, {Prakash}, {Raichoor},
  {Reid}, {Rich}, {Ridl}, {Rodriguez-Torres}, {Carnero Rosell}, {Ross},
  {Rossi}, {Ruan}, {Salvato}, {Sayres}, {Schneider}, {Schlegel}, {Seljak},
  {Seo}, {Sesar}, {Shandera}, {Shu}, {Slosar}, {Sobreira}, {Streblyanska},
  {Suzuki}, {Taylor}, {Tao}, {Tinker}, {Tojeiro}, {Vargas-Maga{\~n}a}, {Wang},
  {Weaver}, {Weinberg}, {White}, {Wood-Vasey}, {Yeche}, {Zhai}, {Zhao}, {Zhao},
  {Zheng}, {Ben Zhu}, \& {Zou}}]{eBOSS}
{Dawson}, K.~S., {Kneib}, J.-P., {Percival}, W.~J., {et~al.} 2016, \aj, 151,
  44, \dodoi{10.3847/0004-6256/151/2/44}

\bibitem[{{Dessauges-Zavadsky} {et~al.}(2017){Dessauges-Zavadsky}, {Zamojski},
  {Rujopakarn}, {Richard}, {Sklias}, {Schaerer}, {Combes}, {Ebeling}, {Rawle},
  {Egami}, {Boone}, {Cl{\'e}ment}, {Kneib}, {Nyland}, \&
  {Walth}}]{Dessauges-Zavadsky17}
{Dessauges-Zavadsky}, M., {Zamojski}, M., {Rujopakarn}, W., {et~al.} 2017,
  \aap, 605, A81, \dodoi{10.1051/0004-6361/201628513}

\bibitem[{{Dey} {et~al.}(2019){Dey}, {Schlegel}, {Lang}, {Blum}, {Burleigh},
  {Fan}, {Findlay}, {Finkbeiner}, {Herrera}, {Juneau}, {Landriau}, {Levi},
  {McGreer}, {Meisner}, {Myers}, {Moustakas}, {Nugent}, {Patej}, {Schlafly},
  {Walker}, {Valdes}, {Weaver}, {Y{\`e}che}, {Zou}, {Zhou}, {Abareshi},
  {Abbott}, {Abolfathi}, {Aguilera}, {Alam}, {Allen}, {Alvarez}, {Annis},
  {Ansarinejad}, {Aubert}, {Beechert}, {Bell}, {BenZvi}, {Beutler}, {Bielby},
  {Bolton}, {Brice{\~n}o}, {Buckley-Geer}, {Butler}, {Calamida}, {Carlberg},
  {Carter}, {Casas}, {Castander}, {Choi}, {Comparat}, {Cukanovaite}, {Delubac},
  {DeVries}, {Dey}, {Dhungana}, {Dickinson}, {Ding}, {Donaldson}, {Duan},
  {Duckworth}, {Eftekharzadeh}, {Eisenstein}, {Etourneau}, {Fagrelius},
  {Farihi}, {Fitzpatrick}, {Font-Ribera}, {Fulmer}, {G{\"a}nsicke},
  {Gaztanaga}, {George}, {Gerdes}, {Gontcho}, {Gorgoni}, {Green}, {Guy},
  {Harmer}, {Hernandez}, {Honscheid}, {Huang}, {James}, {Jannuzi}, {Jiang},
  {Joyce}, {Karcher}, {Karkar}, {Kehoe}, {Kneib}, {Kueter-Young}, {Lan},
  {Lauer}, {Le Guillou}, {Le Van Suu}, {Lee}, {Lesser}, {Perreault Levasseur},
  {Li}, {Mann}, {Marshall}, {Mart{\'\i}nez-V{\'a}zquez}, {Martini}, {du Mas des
  Bourboux}, {McManus}, {Meier}, {M{\'e}nard}, {Metcalfe},
  {Mu{\~n}oz-Guti{\'e}rrez}, {Najita}, {Napier}, {Narayan}, {Newman}, {Nie},
  {Nord}, {Norman}, {Olsen}, {Paat}, {Palanque-Delabrouille}, {Peng},
  {Poppett}, {Poremba}, {Prakash}, {Rabinowitz}, {Raichoor}, {Rezaie},
  {Robertson}, {Roe}, {Ross}, {Ross}, {Rudnick}, {Safonova}, {Saha},
  {S{\'a}nchez}, {Savary}, {Schweiker}, {Scott}, {Seo}, {Shan}, {Silva},
  {Slepian}, {Soto}, {Sprayberry}, {Staten}, {Stillman}, {Stupak}, {Summers},
  {Sien Tie}, {Tirado}, {Vargas-Maga{\~n}a}, {Vivas}, {Wechsler}, {Williams},
  {Yang}, {Yang}, {Yapici}, {Zaritsky}, {Zenteno}, {Zhang}, {Zhang}, {Zhou}, \&
  {Zhou}}]{Dey.etal.2019}
{Dey}, A., {Schlegel}, D.~J., {Lang}, D., {et~al.} 2019, \aj, 157, 168,
  \dodoi{10.3847/1538-3881/ab089d}

\bibitem[{{Dressler} {et~al.}(2011){Dressler}, {Bigelow}, {Hare}, {Sutin},
  {Thompson}, {Burley}, {Epps}, {Oemler}, {Bagish}, {Birk}, {Clardy},
  {Gunnels}, {Kelson}, {Shectman}, \& {Osip}}]{Alan2011}
{Dressler}, A., {Bigelow}, B., {Hare}, T., {et~al.} 2011, \pasp, 123, 288,
  \dodoi{10.1086/658908}

\bibitem[{{El{\'\i}asd{\'o}ttir} {et~al.}(2007){El{\'\i}asd{\'o}ttir},
  {Limousin}, {Richard}, {Hjorth}, {Kneib}, {Natarajan}, {Pedersen}, {Jullo},
  \& {Paraficz}}]{Eliasdottir.etal.2007.pdie}
{El{\'\i}asd{\'o}ttir}, {\'A}., {Limousin}, M., {Richard}, J., {et~al.} 2007,
  arXiv e-prints, arXiv:0710.5636.
\newblock \doarXiv{0710.5636}

\bibitem[{{Furtak} {et~al.}(2022){Furtak}, {Plat}, {Zitrin}, {Topping},
  {Stark}, {Strait}, {Charlot}, {Coe}, {Andrade-Santos}, {Brada{\v{c}}},
  {Bradley}, {Lemaux}, \& {Sharon}}]{Furtak.etal.2022}
{Furtak}, L.~J., {Plat}, A., {Zitrin}, A., {et~al.} 2022, arXiv e-prints,
  arXiv:2204.09668.
\newblock \doarXiv{2204.09668}

\bibitem[{Garg {et~al.}(2006)Garg, Stubbs, Challis, Wood-Vasey, Blondin, Huber,
  Cook, Nikolaev, Rest, Smith, Olsen, Suntzeff, Aguilera, Prieto, Becker,
  Miceli, Miknaitis, Clocchiatti, Minniti, Morelli, \& Welch}]{Garg.etal.2006}
Garg, A., Stubbs, C.~W., Challis, P., {et~al.} 2006, The Astronomical Journal,
  133, 403, \dodoi{10.1086/510118}

\bibitem[{{Gazagnes} {et~al.}(2020){Gazagnes}, {Chisholm}, {Schaerer},
  {Verhamme}, \& {Izotov}}]{Gazagnes.etal.2020}
{Gazagnes}, S., {Chisholm}, J., {Schaerer}, D., {Verhamme}, A., \& {Izotov}, Y.
  2020, \aap, 639, A85, \dodoi{10.1051/0004-6361/202038096}

\bibitem[{{Giavalisco}(2002)}]{Giavalisco.etal.2002}
{Giavalisco}, M. 2002, \araa, 40, 579,
  \dodoi{10.1146/annurev.astro.40.121301.111837}

\bibitem[{{Giavalisco} {et~al.}(1996){Giavalisco}, {Koratkar}, \&
  {Calzetti}}]{giavalisco1996}
{Giavalisco}, M., {Koratkar}, A., \& {Calzetti}, D. 1996, \apj, 466, 831,
  \dodoi{10.1086/177557}

\bibitem[{{Gilbank} {et~al.}(2008){Gilbank}, {Yee}, {Ellingson}, {Hicks},
  {Gladders}, {Barrientos}, \& {Keeney}}]{gilbank08}
{Gilbank}, D.~G., {Yee}, H.~K.~C., {Ellingson}, E., {et~al.} 2008, \apjl, 677,
  L89, \dodoi{10.1086/588138}

\bibitem[{{Haardt} \& {Madau}(1996)}]{Haardt.Madau.1996}
{Haardt}, F., \& {Madau}, P. 1996, \apj, 461, 20, \dodoi{10.1086/177035}

\bibitem[{Harris {et~al.}(2020)Harris, Millman, van~der Walt, Gommers,
  Virtanen, Cournapeau, Wieser, Taylor, Berg, Smith, Kern, Picus, Hoyer, van
  Kerkwijk, Brett, Haldane, del R{\'{\i}}o, Wiebe, Peterson,
  G{\'{e}}rard-Marchant, Sheppard, Reddy, Weckesser, Abbasi, Gohlke, \&
  Oliphant}]{numpy_paper}
Harris, C.~R., Millman, K.~J., van~der Walt, S.~J., {et~al.} 2020, Nature, 585,
  357, \dodoi{10.1038/s41586-020-2649-2}

\bibitem[{Hunter(2007)}]{matplotlib_paper}
Hunter, J.~D. 2007, Computing in Science Engineering, 9, 90,
  \dodoi{10.1109/MCSE.2007.55}

\bibitem[{{Izotov} {et~al.}(2021){Izotov}, {Worseck}, {Schaerer}, {Guseva},
  {Chisholm}, {Thuan}, {Fricke}, \& {Verhamme}}]{Izotov.etal.2021}
{Izotov}, Y.~I., {Worseck}, G., {Schaerer}, D., {et~al.} 2021, \mnras, 503,
  1734, \dodoi{10.1093/mnras/stab612}

\bibitem[{{Johnson}(2019)}]{SEDpy}
{Johnson}, B.~D. 2019, {SEDPY: Modules for storing and operating on
  astronomical source spectral energy distribution}, Astrophysics Source Code
  Library, record ascl:1905.026.
\newblock \doeprint{1905.026}

\bibitem[{{Johnson} {et~al.}(2021){Johnson}, {Leja}, {Conroy}, \&
  {Speagle}}]{Johnson.etal.2021.Prospector}
{Johnson}, B.~D., {Leja}, J., {Conroy}, C., \& {Speagle}, J.~S. 2021, \apjs,
  254, 22, \dodoi{10.3847/1538-4365/abef67}

\bibitem[{{Johnson} {et~al.}(2017){Johnson}, {Rigby}, {Sharon}, {Gladders},
  {Florian}, {Bayliss}, {Wuyts}, {Whitaker}, {Livermore}, \&
  {Murray}}]{johnson2017}
{Johnson}, T.~L., {Rigby}, J.~R., {Sharon}, K., {et~al.} 2017, \apjl, 843, L21,
  \dodoi{10.3847/2041-8213/aa7516}

\bibitem[{{Jones} {et~al.}(2010){Jones}, {Swinbank}, {Ellis}, {Richard}, \&
  {Stark}}]{jones10}
{Jones}, T.~A., {Swinbank}, A.~M., {Ellis}, R.~S., {Richard}, J., \& {Stark},
  D.~P. 2010, \mnras, 404, 1247, \dodoi{10.1111/j.1365-2966.2010.16378.x}

\bibitem[{{Jullo} {et~al.}(2007){Jullo}, {Kneib}, {Limousin},
  {El{\'\i}asd{\'o}ttir}, {Marshall}, \& {Verdugo}}]{Jullo.etal.2007.lenstool}
{Jullo}, E., {Kneib}, J.~P., {Limousin}, M., {et~al.} 2007, New Journal of
  Physics, 9, 447, \dodoi{10.1088/1367-2630/9/12/447}

\bibitem[{{Kennicutt}(1998)}]{Kennicutt.etal.1998}
{Kennicutt}, Robert~C., J. 1998, \araa, 36, 189,
  \dodoi{10.1146/annurev.astro.36.1.189}

\bibitem[{{Khullar} {et~al.}(2021){Khullar}, {Gozman}, {Lin}, {Martinez},
  {Matthews Acu{\~n}a}, {Medina}, {Merz}, {Sanchez}, {Sisco}, {Kavin Stein},
  {Sukay}, {Tavangar}, {Bayliss}, {Bleem}, {Brownsberger}, {Dahle}, {Florian},
  {Gladders}, {Mahler}, {Rigby}, {Sharon}, \& {Stark}}]{Khullar.etal.2021}
{Khullar}, G., {Gozman}, K., {Lin}, J.~J., {et~al.} 2021, \apj, 906, 107,
  \dodoi{10.3847/1538-4357/abcb86}

\bibitem[{{Kroupa}(2001)}]{Kroupa.etal.2001}
{Kroupa}, P. 2001, \mnras, 322, 231, \dodoi{10.1046/j.1365-8711.2001.04022.x}

\bibitem[{{Lang} {et~al.}(2010){Lang}, {Hogg}, {Mierle}, {Blanton}, \&
  {Roweis}}]{astrometry.net}
{Lang}, D., {Hogg}, D.~W., {Mierle}, K., {Blanton}, M., \& {Roweis}, S. 2010,
  \aj, 139, 1782, \dodoi{10.1088/0004-6256/139/5/1782}

\bibitem[{{Lee} {et~al.}(2012){Lee}, {Ferguson}, {Wiklind}, {Dahlen},
  {Dickinson}, {Giavalisco}, {Grogin}, {Papovich}, {Messias}, {Guo}, \&
  {Lin}}]{Lee.etal.2012}
{Lee}, K.-S., {Ferguson}, H.~C., {Wiklind}, T., {et~al.} 2012, \apj, 752, 66,
  \dodoi{10.1088/0004-637X/752/1/66}

\bibitem[{{Leja} {et~al.}(2017){Leja}, {Johnson}, {Conroy}, {van Dokkum}, \&
  {Byler}}]{Leja_2017}
{Leja}, J., {Johnson}, B.~D., {Conroy}, C., {van Dokkum}, P.~G., \& {Byler}, N.
  2017, \apj, 837, 170, \dodoi{10.3847/1538-4357/aa5ffe}

\bibitem[{{Limousin} {et~al.}(2005){Limousin}, {Kneib}, \&
  {Natarajan}}]{Limousin.etal.2005}
{Limousin}, M., {Kneib}, J.-P., \& {Natarajan}, P. 2005, \mnras, 356, 309,
  \dodoi{10.1111/j.1365-2966.2004.08449.x}

\bibitem[{{Lotz} {et~al.}(2017){Lotz}, {Koekemoer}, {Coe}, {Grogin}, {Capak},
  {Mack}, {Anderson}, {Avila}, {Barker}, {Borncamp}, {Brammer}, {Durbin},
  {Gunning}, {Hilbert}, {Jenkner}, {Khandrika}, {Levay}, {Lucas}, {MacKenty},
  {Ogaz}, {Porterfield}, {Reid}, {Robberto}, {Royle}, {Smith},
  {Storrie-Lombardi}, {Sunnquist}, {Surace}, {Taylor}, {Williams}, {Bullock},
  {Dickinson}, {Finkelstein}, {Natarajan}, {Richard}, {Robertson}, {Tumlinson},
  {Zitrin}, {Flanagan}, {Sembach}, {Soifer}, \& {Mountain}}]{Lotz.etal.2017}
{Lotz}, J.~M., {Koekemoer}, A., {Coe}, D., {et~al.} 2017, \apj, 837, 97,
  \dodoi{10.3847/1538-4357/837/1/97}

\bibitem[{{Madau}(1995)}]{Madau.etal.1995}
{Madau}, P. 1995, \apj, 441, 18, \dodoi{10.1086/175332}

\bibitem[{{Madau} \& {Dickinson}(2014)}]{madaureview2014}
{Madau}, P., \& {Dickinson}, M. 2014, \araa, 52, 415,
  \dodoi{10.1146/annurev-astro-081811-125615}

\bibitem[{{Mahler} {et~al.}(2020){Mahler}, {Sharon}, {Gladders}, {Bleem},
  {Bayliss}, {Calzadilla}, {Floyd}, {Khullar}, {McDonald}, {Remolina
  Gonz{\'a}lez}, {Schrabback}, {Stark}, \& {van den Busch}}]{Mahler.etal.2020}
{Mahler}, G., {Sharon}, K., {Gladders}, M.~D., {et~al.} 2020, \apj, 894, 150,
  \dodoi{10.3847/1538-4357/ab886b}

\bibitem[{{Mahler} {et~al.}(2022){Mahler}, {Jauzac}, {Richard}, {Beauchesne},
  {Ebeling}, {Lagattuta}, {Natarajan}, {Sharon}, {Atek}, {Claeyssens},
  {Cl{\'e}ment}, {Eckert}, {Edge}, {Kneib}, \&
  {Niemiec}}]{Mahler.etal.2022.jwstsmacs}
{Mahler}, G., {Jauzac}, M., {Richard}, J., {et~al.} 2022, arXiv e-prints,
  arXiv:2207.07101.
\newblock \doarXiv{2207.07101}

\bibitem[{{Maraston} {et~al.}(2010){Maraston}, {Pforr}, {Renzini}, {Daddi},
  {Dickinson}, {Cimatti}, \& {Tonini}}]{Maraston.etal.2010}
{Maraston}, C., {Pforr}, J., {Renzini}, A., {et~al.} 2010, \mnras, 407, 830,
  \dodoi{10.1111/j.1365-2966.2010.16973.x}

\bibitem[{{Marchesini} {et~al.}(2009){Marchesini}, {van Dokkum}, {F{\"o}rster
  Schreiber}, {Franx}, {Labb{\'e}}, \& {Wuyts}}]{Marchesini.etal.2009}
{Marchesini}, D., {van Dokkum}, P.~G., {F{\"o}rster Schreiber}, N.~M., {et~al.}
  2009, \apj, 701, 1765, \dodoi{10.1088/0004-637X/701/2/1765}

\bibitem[{{Marsan} {et~al.}(2022){Marsan}, {Muzzin}, {Marchesini}, {Stefanon},
  {Martis}, {Annunziatella}, {Chan}, {Cooper}, {Forrest}, {Gomez},
  {McConachie}, \& {Wilson}}]{Marsan.etal.2022}
{Marsan}, Z.~C., {Muzzin}, A., {Marchesini}, D., {et~al.} 2022, \apj, 924, 25,
  \dodoi{10.3847/1538-4357/ac312a}

\bibitem[{{Martinez} {et~al.}(2022){Martinez}, {Napier}, {Cloonan}, {Sukay},
  {Gozman}, {Merz}, {Khullar}, {Lin}, {Matthews Acu{\~n}a}, {Medina},
  {Sanchez}, {Sisco}, {Kavin Stein}, {Tavangar}, {Remolina Gonz{\`a}lez},
  {Mahler}, {Sharon}, {Dahle}, \& {Gladders}}]{COOLLAMPS3}
{Martinez}, M.~N., {Napier}, K.~A., {Cloonan}, A.~P., {et~al.} 2022, arXiv
  e-prints, arXiv:2209.03972.
\newblock \doarXiv{2209.03972}

\bibitem[{{Matthee} \& {Schaye}(2019)}]{Matthee.etal.2019}
{Matthee}, J., \& {Schaye}, J. 2019, \mnras, 484, 915,
  \dodoi{10.1093/mnras/stz030}

\bibitem[{Miknaitis {et~al.}(2007)Miknaitis, Pignata, Rest, Wood-Vasey,
  Blondin, Challis, Smith, Stubbs, Suntzeff, Foley, Matheson, Tonry, Aguilera,
  Blackman, Becker, Clocchiatti, Covarrubias, Davis, Filippenko, Garg,
  Garnavich, Hicken, Jha, Krisciunas, Kirshner, Leibundgut, Li, Miceli,
  Narayan, Prieto, Riess, Salvo, Schmidt, Sollerman, Spyromilio, \&
  Zenteno}]{Miknaitis.etal.2007}
Miknaitis, G., Pignata, G., Rest, A., {et~al.} 2007, The Astrophysical Journal,
  666, 674, \dodoi{10.1086/519986}

\bibitem[{{Muzzin} {et~al.}(2013){Muzzin}, {Marchesini}, {Stefanon}, {Franx},
  {McCracken}, {Milvang-Jensen}, {Dunlop}, {Fynbo}, {Brammer}, {Labb{\'e}}, \&
  {van Dokkum}}]{Muzzin.etal.2013}
{Muzzin}, A., {Marchesini}, D., {Stefanon}, M., {et~al.} 2013, \apj, 777, 18,
  \dodoi{10.1088/0004-637X/777/1/18}

\bibitem[{{Oguri} {et~al.}(2012){Oguri}, {Bayliss}, {Dahle}, {Sharon},
  {Gladders}, {Natarajan}, {Hennawi}, \& {Koester}}]{Oguri.etal.2012.sgas1050}
{Oguri}, M., {Bayliss}, M.~B., {Dahle}, H., {et~al.} 2012, \mnras, 420, 3213,
  \dodoi{10.1111/j.1365-2966.2011.20248.x}

\bibitem[{{Peng} {et~al.}(2010){Peng}, {Ho}, {Impey}, \&
  {Rix}}]{Peng.etal.2010}
{Peng}, C.~Y., {Ho}, L.~C., {Impey}, C.~D., \& {Rix}, H.-W. 2010, \aj, 139,
  2097, \dodoi{10.1088/0004-6256/139/6/2097}

\bibitem[{{Planck Collaboration} {et~al.}(2020){Planck Collaboration},
  {Aghanim}, {Akrami}, {Ashdown}, {Aumont}, {Baccigalupi}, {Ballardini},
  {Banday}, {Barreiro}, {Bartolo}, {Basak}, {Battye}, {Benabed}, {Bernard},
  {Bersanelli}, {Bielewicz}, {Bock}, {Bond}, {Borrill}, {Bouchet}, {Boulanger},
  {Bucher}, {Burigana}, {Butler}, {Calabrese}, {Cardoso}, {Carron},
  {Challinor}, {Chiang}, {Chluba}, {Colombo}, {Combet}, {Contreras}, {Crill},
  {Cuttaia}, {de Bernardis}, {de Zotti}, {Delabrouille}, {Delouis}, {Di
  Valentino}, {Diego}, {Dor{\'e}}, {Douspis}, {Ducout}, {Dupac}, {Dusini},
  {Efstathiou}, {Elsner}, {En{\ss}lin}, {Eriksen}, {Fantaye}, {Farhang},
  {Fergusson}, {Fernandez-Cobos}, {Finelli}, {Forastieri}, {Frailis},
  {Fraisse}, {Franceschi}, {Frolov}, {Galeotta}, {Galli}, {Ganga},
  {G{\'e}nova-Santos}, {Gerbino}, {Ghosh}, {Gonz{\'a}lez-Nuevo}, {G{\'o}rski},
  {Gratton}, {Gruppuso}, {Gudmundsson}, {Hamann}, {Handley}, {Hansen},
  {Herranz}, {Hildebrandt}, {Hivon}, {Huang}, {Jaffe}, {Jones}, {Karakci},
  {Keih{\"a}nen}, {Keskitalo}, {Kiiveri}, {Kim}, {Kisner}, {Knox},
  {Krachmalnicoff}, {Kunz}, {Kurki-Suonio}, {Lagache}, {Lamarre}, {Lasenby},
  {Lattanzi}, {Lawrence}, {Le Jeune}, {Lemos}, {Lesgourgues}, {Levrier},
  {Lewis}, {Liguori}, {Lilje}, {Lilley}, {Lindholm}, {L{\'o}pez-Caniego},
  {Lubin}, {Ma}, {Mac{\'\i}as-P{\'e}rez}, {Maggio}, {Maino}, {Mandolesi},
  {Mangilli}, {Marcos-Caballero}, {Maris}, {Martin}, {Martinelli},
  {Mart{\'\i}nez-Gonz{\'a}lez}, {Matarrese}, {Mauri}, {McEwen}, {Meinhold},
  {Melchiorri}, {Mennella}, {Migliaccio}, {Millea}, {Mitra},
  {Miville-Desch{\^e}nes}, {Molinari}, {Montier}, {Morgante}, {Moss}, {Natoli},
  {N{\o}rgaard-Nielsen}, {Pagano}, {Paoletti}, {Partridge}, {Patanchon},
  {Peiris}, {Perrotta}, {Pettorino}, {Piacentini}, {Polastri}, {Polenta},
  {Puget}, {Rachen}, {Reinecke}, {Remazeilles}, {Renzi}, {Rocha}, {Rosset},
  {Roudier}, {Rubi{\~n}o-Mart{\'\i}n}, {Ruiz-Granados}, {Salvati}, {Sandri},
  {Savelainen}, {Scott}, {Shellard}, {Sirignano}, {Sirri}, {Spencer},
  {Sunyaev}, {Suur-Uski}, {Tauber}, {Tavagnacco}, {Tenti}, {Toffolatti},
  {Tomasi}, {Trombetti}, {Valenziano}, {Valiviita}, {Van Tent}, {Vibert},
  {Vielva}, {Villa}, {Vittorio}, {Wandelt}, {Wehus}, {White}, {White},
  {Zacchei}, \& {Zonca}}]{Planck.etal.2020}
{Planck Collaboration}, {Aghanim}, N., {Akrami}, Y., {et~al.} 2020, \aap, 641,
  A6, \dodoi{10.1051/0004-6361/201833910}

\bibitem[{{Quider} {et~al.}(2010){Quider}, {Shapley}, {Pettini}, {Steidel}, \&
  {Stark}}]{quider10}
{Quider}, A.~M., {Shapley}, A.~E., {Pettini}, M., {Steidel}, C.~C., \& {Stark},
  D.~P. 2010, \mnras, 402, 1467, \dodoi{10.1111/j.1365-2966.2009.16005.x}

\bibitem[{Rest {et~al.}(2005)Rest, Stubbs, Becker, Miknaitis, Miceli,
  Covarrubias, Hawley, Smith, Suntzeff, Olsen, Prieto, Hiriart, Welch, Cook,
  Nikolaev, Huber, Prochtor, Clocchiatti, Minniti, Garg, Challis, Keller, \&
  Schmidt}]{Rest.etal.2005}
Rest, A., Stubbs, C., Becker, A.~C., {et~al.} 2005, The Astrophysical Journal,
  634, 1103, \dodoi{10.1086/497060}

\bibitem[{{Richard} {et~al.}(2021){Richard}, {Claeyssens}, {Lagattuta},
  {Guaita}, {Bauer}, {Pello}, {Carton}, {Bacon}, {Soucail}, {Lyon}, {Kneib},
  {Mahler}, {Cl{\'e}ment}, {Mercier}, {Variu}, {Tamone}, {Ebeling}, {Schmidt},
  {Nanayakkara}, {Maseda}, {Weilbacher}, {Bouch{\'e}}, {Bouwens}, {Wisotzki},
  {de la Vieuville}, {Martinez}, \& {Patr{\'\i}cio}}]{Richard.etal.2021}
{Richard}, J., {Claeyssens}, A., {Lagattuta}, D., {et~al.} 2021, \aap, 646,
  A83, \dodoi{10.1051/0004-6361/202039462}

\bibitem[{{Rigby} {et~al.}(2018){Rigby}, {Bayliss}, {Sharon}, {Gladders},
  {Chisholm}, {Dahle}, {Johnson}, {Paterno-Mahler}, {Wuyts}, \&
  {Kelson}}]{Rigby.etal.2018.megasaura}
{Rigby}, J.~R., {Bayliss}, M.~B., {Sharon}, K., {et~al.} 2018, \aj, 155, 104,
  \dodoi{10.3847/1538-3881/aaa2ff}

\bibitem[{{Rinaldi} {et~al.}(2022){Rinaldi}, {Caputi}, {van Mierlo}, {Ashby},
  {Caminha}, \& {Iani}}]{inaldi.etal.2022}
{Rinaldi}, P., {Caputi}, K.~I., {van Mierlo}, S.~E., {et~al.} 2022, \apj, 930,
  128, \dodoi{10.3847/1538-4357/ac5d39}

\bibitem[{{Rivera-Thorsen} {et~al.}(2017){Rivera-Thorsen}, {Dahle}, {Gronke},
  {Bayliss}, {Rigby}, {Simcoe}, {Bordoloi}, {Turner}, \&
  {Furesz}}]{Rivera-Thorsen.etal.2017}
{Rivera-Thorsen}, T.~E., {Dahle}, H., {Gronke}, M., {et~al.} 2017, \aap, 608,
  L4, \dodoi{10.1051/0004-6361/201732173}

\bibitem[{{Rivera-Thorsen} {et~al.}(2019){Rivera-Thorsen}, {Dahle}, {Chisholm},
  {Florian}, {Gronke}, {Rigby}, {Gladders}, {Mahler}, {Sharon}, \&
  {Bayliss}}]{Rivera.etal.2019.sunburst}
{Rivera-Thorsen}, T.~E., {Dahle}, H., {Chisholm}, J., {et~al.} 2019, Science,
  366, 738, \dodoi{10.1126/science.aaw0978}

\bibitem[{{Salim} {et~al.}(2007{\natexlab{a}}){Salim}, {Rich}, {Charlot},
  {Brinchmann}, {Johnson}, {Schiminovich}, {Seibert}, {Mallery}, {Heckman},
  {Forster}, {Friedman}, {Martin}, {Morrissey}, {Neff}, {Small}, {Wyder},
  {Bianchi}, {Donas}, {Lee}, {Madore}, {Milliard}, {Szalay}, {Welsh}, \&
  {Yi}}]{Salim.etal.2007}
{Salim}, S., {Rich}, R.~M., {Charlot}, S., {et~al.} 2007{\natexlab{a}}, \apjs,
  173, 267, \dodoi{10.1086/519218}

\bibitem[{{Salim} {et~al.}(2007{\natexlab{b}}){Salim}, {Rich}, {Charlot},
  {Brinchmann}, {Johnson}, {Schiminovich}, {Seibert}, {Mallery}, {Heckman},
  {Forster}, {Friedman}, {Martin}, {Morrissey}, {Neff}, {Small}, {Wyder},
  {Bianchi}, {Donas}, {Lee}, {Madore}, {Milliard}, {Szalay}, {Welsh}, \&
  {Yi}}]{salim2007}
---. 2007{\natexlab{b}}, \apjs, 173, 267, \dodoi{10.1086/519218}

\bibitem[{{Salmon} {et~al.}(2015){Salmon}, {Papovich}, {Finkelstein}, {Tilvi},
  {Finlator}, {Behroozi}, {Dahlen}, {Dav{\'e}}, {Dekel}, {Dickinson},
  {Ferguson}, {Giavalisco}, {Long}, {Lu}, {Mobasher}, {Reddy}, {Somerville}, \&
  {Wechsler}}]{Salmon.etal.2015}
{Salmon}, B., {Papovich}, C., {Finkelstein}, S.~L., {et~al.} 2015, \apj, 799,
  183, \dodoi{10.1088/0004-637X/799/2/183}

\bibitem[{{Santini} {et~al.}(2009){Santini}, {Fontana}, {Grazian}, {Salimbeni},
  {Fiore}, {Fontanot}, {Boutsia}, {Castellano}, {Cristiani}, {de Santis},
  {Gallozzi}, {Giallongo}, {Menci}, {Nonino}, {Paris}, {Pentericci}, \&
  {Vanzella}}]{Santini.etal.2009}
{Santini}, P., {Fontana}, A., {Grazian}, A., {et~al.} 2009, \aap, 504, 751,
  \dodoi{10.1051/0004-6361/200811434}

\bibitem[{{Santini} {et~al.}(2017){Santini}, {Fontana}, {Castellano}, {Di
  Criscienzo}, {Merlin}, {Amorin}, {Cullen}, {Daddi}, {Dickinson}, {Dunlop},
  {Grazian}, {Lamastra}, {McLure}, {Micha{\l}owski}, {Pentericci}, \&
  {Shu}}]{Santini.etal.2017}
{Santini}, P., {Fontana}, A., {Castellano}, M., {et~al.} 2017, \apj, 847, 76,
  \dodoi{10.3847/1538-4357/aa8874}

\bibitem[{{Schlafly} \& {Finkbeiner}(2011)}]{MKExt}
{Schlafly}, E.~F., \& {Finkbeiner}, D.~P. 2011, \apj, 737, 103,
  \dodoi{10.1088/0004-637X/737/2/103}

\bibitem[{{S{\'e}rsic}(1963)}]{Sersic_1963}
{S{\'e}rsic}, J.~L. 1963, Boletin de la Asociaci\'on Argentina de Astronom\'ia
  La Plata Argentina, 6, 41

\bibitem[{{Sharon} {et~al.}(2020){Sharon}, {Bayliss}, {Dahle}, {Dunham},
  {Florian}, {Gladders}, {Johnson}, {Mahler}, {Paterno-Mahler}, {Rigby},
  {Whitaker}, {Akhshik}, {Koester}, {Murray}, {Remolina Gonz{\'a}lez}, \&
  {Wuyts}}]{Sharon.etal.2020}
{Sharon}, K., {Bayliss}, M.~B., {Dahle}, H., {et~al.} 2020, \apjs, 247, 12,
  \dodoi{10.3847/1538-4365/ab5f13}

\bibitem[{{Shivaei} {et~al.}(2015){Shivaei}, {Reddy}, {Steidel}, \&
  {Shapley}}]{shivaei2015}
{Shivaei}, I., {Reddy}, N.~A., {Steidel}, C.~C., \& {Shapley}, A.~E. 2015,
  \apj, 804, 149, \dodoi{10.1088/0004-637X/804/2/149}

\bibitem[{{Simcoe} {et~al.}(2013){Simcoe}, {Burgasser}, {Schechter}, {Fishner},
  {Bernstein}, {Bigelow}, {Pipher}, {Forrest}, {McMurtry}, {Smith}, \&
  {Bochanski}}]{simcoe2013}
{Simcoe}, R.~A., {Burgasser}, A.~J., {Schechter}, P.~L., {et~al.} 2013, \pasp,
  125, 270, \dodoi{10.1086/670241}

\bibitem[{{Smail} {et~al.}(2007){Smail}, {Swinbank}, {Richard}, {Ebeling},
  {Kneib}, {Edge}, {Stark}, {Ellis}, {Dye}, {Smith}, \&
  {Mullis}}]{Smail.etal.2007.cosmiceye}
{Smail}, I., {Swinbank}, A.~M., {Richard}, J., {et~al.} 2007, \apjl, 654, L33,
  \dodoi{10.1086/510902}

\bibitem[{{Sobral} \& {Matthee}(2019)}]{sobral2019}
{Sobral}, D., \& {Matthee}, J. 2019, \aap, 623, A157,
  \dodoi{10.1051/0004-6361/201833075}

\bibitem[{{Stark} {et~al.}(2013){Stark}, {Auger}, {Belokurov}, {Jones},
  {Robertson}, {Ellis}, {Sand}, {Moiseev}, {Eagle}, \&
  {Myers}}]{Stark.etal.2013.cassowary}
{Stark}, D.~P., {Auger}, M., {Belokurov}, V., {et~al.} 2013, \mnras, 436, 1040,
  \dodoi{10.1093/mnras/stt1624}

\bibitem[{{Sukay} {et~al.}(2022){Sukay}, {Khullar}, {Gladders}, {Sharon},
  {Mahler}, {Napier}, {Bleem}, {Dahle}, {Florian}, {Gozman}, {Lin}, {Martinez},
  {Matthews Acu{\~n}a}, {Medina}, {Merz}, {Sanchez}, {Sisco}, {Kavin Stein},
  {Tavangar}, \& {Whitaker}}]{Sukay.etal.2022}
{Sukay}, E., {Khullar}, G., {Gladders}, M.~D., {et~al.} 2022, arXiv e-prints,
  arXiv:2203.11957.
\newblock \doarXiv{2203.11957}

\bibitem[{{Tasca} {et~al.}(2015){Tasca}, {Le F{\`e}vre}, {Hathi}, {Schaerer},
  {Ilbert}, {Zamorani}, {Lemaux}, {Cassata}, {Garilli}, {Le Brun}, {Maccagni},
  {Pentericci}, {Thomas}, {Vanzella}, {Zucca}, {Amorin}, {Bardelli},
  {Cassar{\`a}}, {Castellano}, {Cimatti}, {Cucciati}, {Durkalec}, {Fontana},
  {Giavalisco}, {Grazian}, {Paltani}, {Ribeiro}, {Scodeggio}, {Sommariva},
  {Talia}, {Tresse}, {Vergani}, {Capak}, {Charlot}, {Contini}, {de la Torre},
  {Dunlop}, {Fotopoulou}, {Koekemoer}, {L{\'o}pez-Sanjuan}, {Mellier}, {Pforr},
  {Salvato}, {Scoville}, {Taniguchi}, \& {Wang}}]{Tasca.etal.2015}
{Tasca}, L.~A.~M., {Le F{\`e}vre}, O., {Hathi}, N.~P., {et~al.} 2015, \aap,
  581, A54, \dodoi{10.1051/0004-6361/201425379}

\bibitem[{Tody(1986)}]{Tody.1986}
Tody, D. 1986, in Instrumentation in Astronomy VI, ed. D.~L. Crawford, Vol.
  0627, International Society for Optics and Photonics (SPIE), 733 -- 748,
  \dodoi{10.1117/12.968154}

\bibitem[{{Tody}(1993)}]{Tody.1993}
{Tody}, D. 1993, in Astronomical Society of the Pacific Conference Series,
  Vol.~52, Astronomical Data Analysis Software and Systems II, ed. R.~J.
  {Hanisch}, R.~J.~V. {Brissenden}, \& J.~{Barnes}, 173

\bibitem[{{Tomczak} {et~al.}(2016){Tomczak}, {Quadri}, {Tran}, {Labb{\'e}},
  {Straatman}, {Papovich}, {Glazebrook}, {Allen}, {Brammer}, {Cowley},
  {Dickinson}, {Elbaz}, {Inami}, {Kacprzak}, {Morrison}, {Nanayakkara},
  {Persson}, {Rees}, {Salmon}, {Schreiber}, {Spitler}, \&
  {Whitaker}}]{Tomczak.etal.2016}
{Tomczak}, A.~R., {Quadri}, R.~F., {Tran}, K.-V.~H., {et~al.} 2016, \apj, 817,
  118, \dodoi{10.3847/0004-637X/817/2/118}

\bibitem[{{Tran} {et~al.}(2022){Tran}, {Harshan}, {Glazebrook}, {Keerthi
  Vasan}, {Jones}, {Jacobs}, {Kacprzak}, {Barone}, {Collett}, {Gupta},
  {Henderson}, {Kewley}, {Lopez}, {Nanayakkara}, {Sanders}, \&
  {Sweet}}]{Tran.etal.2022}
{Tran}, K.-V.~H., {Harshan}, A., {Glazebrook}, K., {et~al.} 2022, \aj, 164,
  148, \dodoi{10.3847/1538-3881/ac7da2}

\bibitem[{{Vanzella} {et~al.}(2018){Vanzella}, {Nonino}, {Cupani},
  {Castellano}, {Sani}, {Mignoli}, {Calura}, {Meneghetti}, {Gilli}, {Comastri},
  {Mercurio}, {Caminha}, {Caputi}, {Rosati}, {Grillo}, {Cristiani}, {Balestra},
  {Fontana}, \& {Giavalisco}}]{Vanzella.etal.2018}
{Vanzella}, E., {Nonino}, M., {Cupani}, G., {et~al.} 2018, \mnras, 476, L15,
  \dodoi{10.1093/mnrasl/sly023}

\bibitem[{Virtanen {et~al.}(2020)Virtanen, Gommers, Oliphant, Haberland, Reddy,
  Cournapeau, Burovski, Peterson, Weckesser, Bright, {van der Walt}, Brett,
  Wilson, Millman, Mayorov, Nelson, Jones, Kern, Larson, Carey, Polat, Feng,
  Moore, {VanderPlas}, Laxalde, Perktold, Cimrman, Henriksen, Quintero, Harris,
  Archibald, Ribeiro, Pedregosa, {van Mulbregt}, \& {SciPy 1.0
  Contributors}}]{2020SciPy-NMeth}
Virtanen, P., Gommers, R., Oliphant, T.~E., {et~al.} 2020, Nature Methods, 17,
  261, \dodoi{10.1038/s41592-019-0686-2}

\bibitem[{{Walcher} {et~al.}(2011){Walcher}, {Groves}, {Budav{\'a}ri}, \&
  {Dale}}]{Walcher.etal.2011}
{Walcher}, J., {Groves}, B., {Budav{\'a}ri}, T., \& {Dale}, D. 2011, \apss,
  331, 1, \dodoi{10.1007/s10509-010-0458-z}

\bibitem[{{Wilkins} {et~al.}(2012){Wilkins}, {Gonzalez-Perez}, {Lacey}, \&
  {Baugh}}]{Wilkins.etal.2012}
{Wilkins}, S.~M., {Gonzalez-Perez}, V., {Lacey}, C.~G., \& {Baugh}, C.~M. 2012,
  \mnras, 427, 1490, \dodoi{10.1111/j.1365-2966.2012.22092.x}

\bibitem[{{Wu} {et~al.}(2022){Wu}, {Cai}, {Sun}, {Bian}, {Lin}, {Li}, {Li},
  {Bauer}, {Egami}, {Fan}, {Gonz{\'a}lez-L{\'o}pez}, {Li}, {Wang}, {Yang},
  {Zhang}, \& {Zou}}]{Wu.etal.2022.jwstlens}
{Wu}, Y., {Cai}, Z., {Sun}, F., {et~al.} 2022, arXiv e-prints,
  arXiv:2208.08473.
\newblock \doarXiv{2208.08473}

\end{thebibliography}

\end{document}